\documentclass[a4paper,11pt]{article}
\pdfoutput=1
\usepackage{jcappub} 

\title{\boldmath Dark energy on astrophysical scales and its detection in the Milky Way}
\author[a]{Rui Zhang}
\author[b,1]{and Zhen Zhang\note{Corresponding author.}}

\affiliation[a]{Theoretical Physics Division, Institute of High Energy Physics, \\
Chinese Academy of Sciences, \\
19B Yuquan Road, Beijing 100049, People's Republic of China}
\affiliation[b]{Key Laboratory of Particle Astrophysics, Institute of High Energy Physics, \\
Chinese Academy of Sciences, \\
19B Yuquan Road, Beijing 100049, People's Republic of China}

\emailAdd{zhangzhen@ihep.ac.cn}

\abstract{
The origin and nature of dark energy is one of the most significant challenges in modern science. This research aims to investigate dark energy on astrophysical scales and provide a cosmology-independent method to measure its equation-of-state parameter $w$. To accomplish this, we introduce the concept of a perfect fluid in any static, curved spacetime, and express the energy-momentum tensor of the perfect fluid in a general isotropic form, namely Weinberg's isotropic form. This enables us to define an equation-of-state parameter in a physical and global manner. Within this theoretical framework, we demonstrate that the energy-momentum tensor of dark energy on different scales can take the general isotropic form. Furthermore, we explore the SdS$_{w}$ spacetime and establish its connection with dark energy in cosmology through the equation-of-state parameter $w$. In the SdS$_{w}$ spacetime, a repulsive dark force can be induced by dark energy locally. We then apply the concept of the dark force to realistic astrophysical systems using the Poisson equation. Finally, we find that an anomaly in the Milky Way rotation curve can be quantitatively interpreted by the dark force. By fitting the galactic curve, we are able to obtain the value of the equation-of-state parameter of dark energy, independently of specific dark energy models. 
}

\begin{document}
\maketitle
\flushbottom


\section{Introduction}
The accelerating cosmic expansion implies that the universe is possibly dominated by some component termed as dark energy (DE)~\cite{Riess:1998cb, Perlmutter:1998np}. 
There are kinds of ways proposed to detect DE~\cite{PhysRevD.76.043006,Ishak:2007ea,Fernando:2012ue,Fernando:2014rsa,He:2017alg,Zhang:2021ygh,Ho:2015nsa,Balaguera-Antolinez:2006nwo,Balaguera-Antolinez:2007csw,Vagnozzi:2021quy,Ferlito:2022mok,Nunes:2022bhn}. 
So far the existing DE detections are mostly cosmology-dependent~\cite{Weinberg:2013raj,Weinberg:2013agg}. 
Actually, DE has remained mysterious in its origin and nature. 
What we know about the DE is that it can be characterized by an equation-of-state (EoS) parameter~\cite{Cai:2009zp,Li:2011sd,Wang:2016och}
\begin{equation}
w\equiv p_w/\rho_w\,, 
\end{equation}
where $\rho_w$ and $p_w$ are the energy density and pressure of DE, respectively.

Until today, various DE models have been proposed to explain astronomical observations. 
Generally, these models can be distinguished by their corresponding values of $w$~\cite{He:2017alg,Zhang:2021ygh}. 
For instance, the cosmological constant model has $w=-1$~\cite{1917SPAW.......142E}, the quintessence model has $-1<w<-\frac{1}{3}$~\cite{Wetterich:1987fm,PhysRevLett.82.896}, and the phantom model has $w<-1$~\cite{Caldwell:1999ew}. 
Therefore, the EoS parameter $w$ can serve as a good probe for identifying the appropriate DE model.

If assuming the DE parameter $w$ to be globally a constant in the universe,
one can obtain $w=-1.028\pm0.031$ from fitting the $w$CDM model to the cosmological data~\cite{Aghanim:2018eyx}. 
However, there is no reason why $w$ should be a constant. 
In fact, \cite{Zhao:2017cud} shows that $w$CDM model is not preferred at a $3.5 \sigma$ significance level. 
This means that $w$ may evolve with the cosmological redshift $z_{r}$, namely $w=w\left(z_{r}\right)$.
Nevertheless, $w$ can be regarded as a constant on astrophysical scales, like the Milky Way (MW) galaxy.
Consequently, in the ideal case of a point-like mass, local DE effects can be described well by the SdS$_{w}$ metric~\cite{He:2017alg}.

\begin{figure}
\centerline{
\includegraphics[width=0.68\columnwidth]{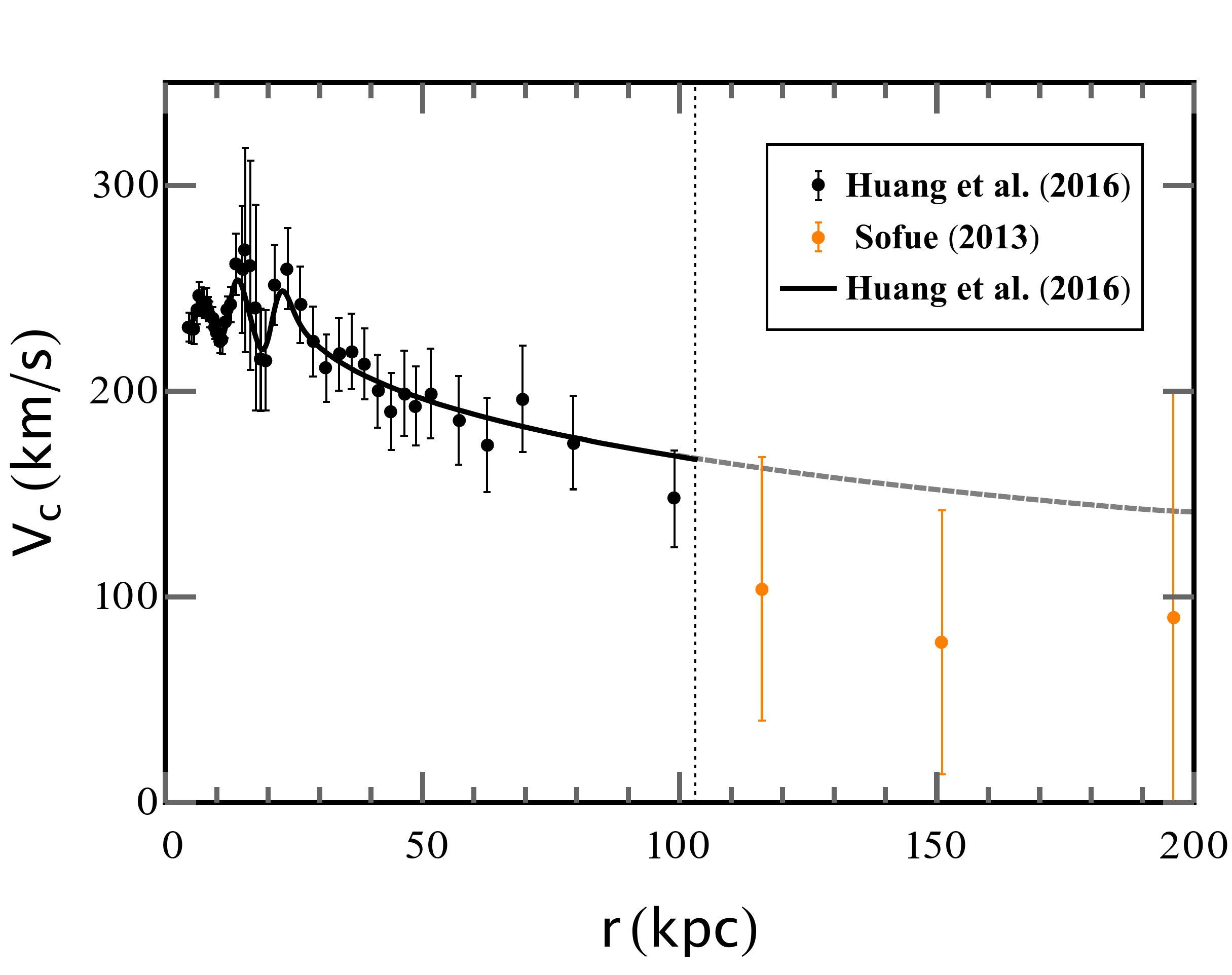}
}
\caption{
The MW rotation curve. The black and orange points are obtained from the Huang et al. 2016 data~\cite{2016MNRAS.463.2623H} and the Sofue 2013 data~\cite{Sofue:2013kja}, respectively. The black line represents the best fitting curve to the Huang et al. 2016 data~\cite{2016MNRAS.463.2623H}. 
We extrapolate the black line to the outer MW region with radius $r\sim100-200$ kpc, where $r$ is the galactocentric radius.
Here we use the gray dashed line to mark this extrapolation. 
Obviously the rotation curve shows an anomalous drop in the outer MW region,
indicating possibly the existence of a repulsive force in that region. 
}
\label{sample}
\end{figure}

The SdS$_{w}$ metric was discovered by Kiselev for the quintessence DE candidate \cite{Kiselev:2002dx}.
Now it has been extended to describe various kinds of DE candidates~\cite{He:2017alg}.
The metric does have many interesting physical and mathematical properties
~\cite{Chen:2005qh,Chen:2008ra,Fernando:2013mex,Azreg-Ainou:2012zyx,Azreg-Ainou:2014lua,Ghosh:2015ovj,Heydarzade:2017wxu,Ghaffarnejad:2018exz}.
Additionally, it may have some practical applications in the DE detections
~\cite{Novosyadlyj:2012qu,Ding:2013vta,Abdujabbarov:2015pqp,Ghosh:2016ddh,Smerechynskyi:2020cfu,Perivolaropoulos:2023zzc}.
For example, a local dark force can be induced by any DE candidate;
in the SdS$_{w}$ spacetime, the specific form of the dark force was derived in a way independent of any specific DE models~\cite{Zhang:2021ygh}. 
In the CC case, under the weak field approximation, it reduces to the one shown in~\cite{Balaguera-Antolinez:2006nwo,Ho:2015nsa}. 
The dark force is locally repulsive, with its strength depending on the DE parameter $w$,
and its existence can lead to deviations from the purely Newtonian force~\cite{Zhang:2021ygh}.
Detecting any deviation from the inverse square law behavior
can help to measure the value of $w$ on astrophysical scales.

Recently, we notice that in the MW galaxy, there seems to be an anomalous drop in the outer rotation curve, 
far more than we expected, as illustrated in Figure~\ref{sample}.
Here, we suggest that this abnormal phenomenon is related with the repulsive dark force that tends to be strengthened greatly in the outer MW region,
although the exact form of the dark force needs further derivations when dealing with a realistic astrophysical system.

In this work, we investigate the dark force acting on astrophysical scales and explore its cosmological origin through the EoS parameter $w$. To gain a comprehensive understanding of the role of $w$ in the dark force and the universe, we extend the Poisson equation to the SdS$_{w}$ spacetime, which enables us to derive the most general gravitational potential that takes into account both matter and DE contributions, and deduce several fundamental cosmological equations, including the generalized Friedmann equations, within an extended Robertson-Walker (RW) spacetime framework. By comparing the EoS parameter in the SdS$_{w}$ spacetime with that in the universe, we establish a connection between DE in cosmology and its astrophysical counterpart. Using the dark force derived from the gradient of the general potential, we propose a novel method for measuring the EoS parameter $w$ by fitting galactic rotation curves. Finally, we summarize our findings and conclusions.


\section{Theoretical analysis}\label{ThA}

\subsection{Cosmological equations}\label{ExtRW} 

The RW metric is used to describe the cosmic expansion. In reality, the universe is not perfectly homogeneous. Thus, the RW metric should be modified to be~\cite{Bardeen:1980kt,Ma:1995ey,Esposito-Farese:2000pbo}
\begin{eqnarray}
\label{eq:ERW}
\begin{array}{rcl}
\displaystyle
\mathrm{d} s^{2}=\big[Z\left(x,y,z\right)\big]^{2}\mathrm{d} t^{2}-\big[a\left(t\right)R\left(x,y,z\right)\big]^{2}\left(\mathrm{d} x^{\,2}+\mathrm{d} y^{\,2}+\mathrm{d} z^{\,2}\right),
\end{array}
\end{eqnarray} 
where $t$ is the time-like coordinate, $a=a\left(t\right)$ is the cosmological expansion factor, as well as $Z\left(x,y,z\right)$ and $R\left(x,y,z\right)$ are functions of the space-like coordinates $x,~y,~z$. The inhomogeneity of the universe may come from the small perturbations excited by primordial quantum fluctuations or catastrophic astrophysical events. 
In the following, this metic with $Z\left(x,y,z\right)\equiv constant$ will be called the {\it extended Robertson-Walker} (ERW) metric,
which can be treated as a background field.

By the law of energy-momentum conservation in an expanding universe, one has 
\begin{align}
\label{eq:EPConservation}
\nabla_{\mu}\,T^{\mu\nu}=\partial_{\mu}\,T^{\mu\nu}+\Gamma^{\mu}_{\,\,\,\mu\rho}\,T^{\rho\nu}+\Gamma^{\nu}_{\,\,\,\mu\rho}\,T^{\mu\rho}=0, 
\end{align}
with $\mu,\,\nu,\,\rho=0,\,1,\,2,\,3$, where $\nabla_{\mu}$ is the covariant derivative, 
$T^{\mu\nu}$ denotes the component of an energy-momentum tensor, and $\Gamma^\mu_{\,\,\,\nu\rho}$ represents the Christoffel symbol.
Considering the zero component of the energy conservation equation, as seen from a comoving observer, it should be written as (appendix~\ref{App:A})
\begin{align}
\label{eq:EConservation}
\frac{\mathrm{d} T^{0}_{\,\,\,0}}{\mathrm{d} t}+3\,\frac{\dot{a}}{a}\left(T^0_{\,\,\,0}-\frac{1}{3}T^{i}_{\,\,\,i}\right)=0, 
\end{align}
with $\dot{a}=\mathrm{d} a/ \mathrm{d} t$, where $T^{\mu}_{\,\,\,\nu}$ represents the mixed component of the energy-momentum tensor. General forms of the conservation equations of energy and momentum can be found in appendix~\ref{App:A}.

From the Einstein equation, 
\begin{align}
G_{\mu\nu}=R_{\mu\nu}-\frac{1}{2}\,g_{\mu\nu}\,R=T_{\mu\nu},
\end{align}
we can derive the generalized Friedmann equations from the ERW metric (appendix~\ref{App:B}): 
\begin{align}
\label{eq:Friedmann2M}
\frac{1}{Z^2}\frac{\ddot{a}}{a}=&-\frac{4}{3}\pi \left(T^{0}_{\,\,\,0}-T^{i}_{\,\,\,i}\right),\\[1.5mm]
\label{eq:Friedmann1M}
\frac{1}{Z^2}\left(\frac{\dot{a}}{a}\right)^2=&-\frac{K\left(x,y,z\right)}{a^2}+\frac{8}{3}\pi \,T^{0}_{\,\,\,0},
\end{align}
where $K=K\left(x,y,z\right)$ is actually a generalization of the constant curvature of the standard RW space.
As demonstrated in appendix~\ref{App:C}, we prove 
\begin{eqnarray}
\label{eq:ERW-Kterm0}
\nonumber
K=\frac{~\sum K^i_{p}~}{~3~}=\frac{1}{3}\,\bigg[-\frac{2}{R^2}\frac{\partial^2_i R}{R}+\frac{1}{R^2}\left(\frac{\partial_i R}{R}\right)^2\bigg], 
\end{eqnarray}
where $K^i_{p}$ is a sectional curvature at the spacetime point $p$. 
Therefore, $K$ is intrinsically an average sectional curvature. 

The expansion rate of the universe is usually characterized by the Hubble parameter,
\begin{align*}
H=\frac{\dot{a}}{a},
\end{align*}
whose present value $H_{0}$ is called the Hubble constant.
Then, the generalized Friedmann equation~\eqref{eq:Friedmann1M} can be reexpressed in a more familiar form,
\begin{align*}
\frac{1}{Z^2}H^2=&-\frac{K\left(x,y,z\right)}{a^2}+\frac{8}{3}\pi \,\rho,
\end{align*}
which takes a much more general form than the traditional Friedmann equation that corresponds to the standard RW metric.

In fluid mechanics, under the incompressibility condition, a fluid can be characterized by the \textit{stress tensor}~\cite{landau2013fluid}, 
\begin{eqnarray}
\label{eq:FM}
\nonumber
T^{i}_{\,\,\,j}\equiv-p\,\delta^{i}_{\,\,\,j}+\tau^{i}_{\,\,\,j},
\end{eqnarray}
where the trace of $\tau^{i}_{\,\,\,j}$ is zero. Then, we find 
\begin{eqnarray}
\label{eq:mpreasure}
\nonumber
\mathrm{Tr}\big[\,T^{i}_{\,\,\,j}\,\big]=-\mathrm{Tr}\big[\,p\,\delta^{i}_{\,\,\,j}\,\big]+\mathrm{Tr}\big[\,\tau^{i}_{\,\,\,j}\,\big]=-3\,p,
\end{eqnarray}
where the trace of the stress tensor is actually equivalent to the mean of the principal stresses of the incompressible fluid~\cite{landau2013fluid}.
In fact, the pressure of any fluid can be defined in a general way as negative one-third of the trace of the stress tensor~\cite{kundu04}, which is sometimes named as the dynamical pressure in fluid mechanics.
Accordingly, we write the energy density and the dynamical pressure for matter or DE in the ERW spacetime as 
\begin{align}
\label{eq:rho-pressure}
\rho\equiv T^{0}_{\,\,\,0}\,, \quad p\equiv-\frac{~\mathrm{Tr}\big[\,T^{i}_{\,\,\,j}\,\big]~}{3}=-\frac{~\sum\,T^{i}_{\,\,\,i}~}{3},
\end{align}
where $\rho$ is the same as the usual energy density, while $p$ differs somewhat from the usual pressure defined in classical fluid mechanics. 
In our case, the energy density defined in~\eqref{eq:rho-pressure} is just the measured one by a local static observer,
and its value does not depend the choice of coordinates. 
Generally, the trace of all the components of the energy-momentum tensor is invariant under coordinate transformations, and thus it is physical.
Once the energy density is given, the dynamical pressure will be independent of specific coordinates and thus physically measurable, 
as a result of its equivalence to the trace of all the spatial components of the energy-momentum tensor or the sum of  $T^{i}_{\,\,\,i}$s that has already been introduced in the conservation equations and the Friedmann equations.

In general relativity, 
both the diagonal and off-diagonal components of the stress tensor are highly dependent of the coordinate system chosen [See, for example, after equation~\eqref{eq:fullT} below]. 
Therefore, the EoS parameter cannot be simply defined as the ratio of some diagonal component of the stress tensor to the energy density.
However, the trace of the stress tensor is a scalar independent of spatial coordinates. 
Accordingly, we obtain an EoS parameter
\begin{align}
\label{eq:EOS}
w=w\left(t, x, y, z\right)=\frac{~p~}{\rho},
\end{align}
which is actually a measurable parameter, independent of the basis referred and the coordinate system chosen.

The specific derivations of the conservation equations and the Friedmann equations, associated with the ERW metric~\eqref{eq:ERW}, 
are presented in appendies~\ref{App:A} and~\ref{App:B}, respectively. 
By carefully analyzing the derivations of the fundamental equations in cosmology,
it is revealed that the off-diagonal components of the energy-momentum tensor do not play any role in the cosmological evolution. 
Indeed, the evolution of the universe is determined by both the $00$ component, i.e., $T^{0}_{\,\,\,0}$, and the trace of the components of the stress tensor, i.e., the average value of the diagonal components, rather than one of these components, which has long been ignored. 
More exactly, it is the EoS parameter, defined in equation~\eqref{eq:EOS} via equation~\eqref{eq:rho-pressure}, that determines the evolution of the universe, as illustrated in equations~\eqref{eq:EConservation}, \eqref{eq:Friedmann2M} and~\eqref{eq:Friedmann1M}.
Generally, the ERW universe contains various forms of components, like DE, matter, and radiation.
The DE component contributes to the total energy density $\rho$~\eqref{eq:rho-pressure}, the total pressure $p$~\eqref{eq:rho-pressure}, and the cosmological EoS parameter $w$~\eqref{eq:EOS} only as a part.  As shown in equation~\eqref{eq:EOS}, there may be a spatial dependence for these three quantities involved in those cosmological equations associated with the ERW metric.
In addition to DE, this dependence may be partially resulted from the contributions of matter and radiation.
Compared with the other two components, DE tends to affect the three quantities on much larger scales.

Based on the ERW metric, it is possible to propose cosmological models 
and test the inhomogeneity of the universe on much larger scales than that of a galaxy, such as that of the large-scale structure. 
But when coming to shorter astrophysical scales, we need to focus on the physics in a realistic astrophysical system like the MW galaxy. 
Compared with the entire universe, a galaxy can only be treated as a ``dust".
Thus we need to develop a cosmology-independent methodology to probe the nearby background structures surrounding the ``dust" at astrophysical scales.
Especially for the DE background, its local properties remain mysterious on such small scales. 
At present, the only knowledge about DE is that its EoS parameter may evolve with the cosmological redshift $z_{r}$, namely, $w=w\left(z_{r}\right)$~\cite{Zhao:2017cud}. 
So far there has not been any method to differentiate the DE background of the ``dust" from the DE counterpart in a standard RW universe.
Conversely, this does indicate that the local DE background can be described by the DE candidate in the standard RW universe to a good approximation.
Therefore, in the following, our analysis will be only performed in the standard RW universe.

In the standard RW universe, the expansion factor evolves with time.
Thus, the EoS parameter may evolve as a function of time. Then, equation~\eqref{eq:EConservation} becomes
\begin{align}
\frac{\dot{\rho}}{\rho}=-3\left(1+w\right)\frac{\dot{a}}{a},
\end{align}
which is derived directly from the ERW metric. If $w$ is a global constant, this equation can be solved to yield 
\begin{align}
\rho\propto a^{-3\,\left(1+w\right)}\,. 
\end{align}
In general, the EoS parameter may evolve with the cosmological redshift $z_{r}$, namely, $w=w\left(z_{r}\right)$. 
However, it takes an approximately constant value on astrophysical scales~\cite{He:2017alg}.

Similarly, the expansion factor $a$ is constant on astrophysical scales, such as that of a galaxy \cite{He:2017alg}.
Thus, it can be absorbed by re-defining the spatial coordinates $\left(x,~y,~z\right)$.
Let us write $x^{\mu}~\left(\mu=0,~1,~2,~3\right)$ for $\left(t,~x,~y,~z\right)$.
Therefore, DE can be thought to be in a locally static state. Accordingly, its energy-momentum tensor can be described by 
\begin{equation}
\label{eq:fullT}
\widetilde{T}=T_{\mu\nu}\,\mathrm{d} x^{\mu}\otimes\mathrm{d} x^{\nu},
\end{equation}
where all the tensor components $T_{\mu\nu}$ are independent of the coordinate time. 
It is perhaps noteworthy that in general relativity, the components of the energy-momentum tensor are closely related with the referred basis $\{\mathrm{d} x^{\mu}\}$.
The referred basis are highly dependent of the coordinate system chosen and hence also both the diagonal and off-diagonal components.
Therefore, judging if a fluid is isotropic based on some of the components of the energy-momentum tensor of the fluid makes no sense.


\subsection{The SdS$_{w}$ spacetime}\label{SdSw}

As mentioned above, the EoS parameter $w$ can be reasonably assumed to be a constant for the DE counterpart in an astrophysical system at redshift $z_{r}$;
its value is inherited directly from that of the cosmological DE at the same redshift. 
Basing on this, we find that, under gravitational fields of matter, the energy-momentum tensor of DE may no longer take the same form as before. 
Taking the quintessence model \cite{Wetterich:1987fm,PhysRevLett.82.896} for example, to keep $w$ constant in an astrophysical system, 
the quintessence DE candidate may redistribute itself on that scale, 
resulting in a deviation of its stress tensor from the traditional isotropic form $T^{i}_{\,\,\,j}\propto\delta^{i}_{\,\,\,j}$.
However, the energy-momentum tensor should obey the Einstein equation.
To obtain an appropriate expression for the energy-momentum tensor of isotropic DE, we need to solve the Einstein equation for an astrophysical system.

In current models, the DE's energy density and pressure are believed to be spatially homogeneous and isotropic on the cosmological scale. 
At much shorter astrophysical scales, the gravitational effects on DE may be significant and should be taken into account as those on matter. 
Especially, in the case of an astrophysical system with a point-like mass, it is likely that the tensor $\widetilde{T}$ can be expressed in a static and spherically-symmetric form.
Denote $\delta_{ij}$ as the Kronecker delta. Then, set $x_{i}=\delta_{ij}\,x^{j}$ and let $\vec{x}\cdot\vec{x}=x_{i}\,x^{i}=r^{2}$. 
By the assumption of staticity and spherical symmetry, the general energy-momentum tensor for any DE candidate can be therefore given in some Cartesian coordinate system by 
\begin{align}
\label{eq:Tcomponets}
\begin{array}{rcl}
\displaystyle
T^t_{\,\,\,t}&=&A(r),\quad T^t_{\,\,\,i}=0,\quad T^i_{\,\,\,j}=B(r)\,\delta^{i}_{\,\,j}+C(r)\,x^{i}\,x_{j},
\end{array}
\end{align}
which was shown by Kiselev in \cite{Kiselev:2002dx},
with the metric
\begin{equation}
\label{eq:sphcartesian}
\mathrm{d} s^{2}=g_{tt}\,\mathrm{d} t^{2}-\bigg[\left(g_{rr}-1\right)\frac{x^{i}\,x^{j}}{r^2}+\delta^{ij}\bigg]\mathrm{d} x_{i}\mathrm{d} x_{j},
\end{equation}
where $A(r)$, $B(r)$, $C(r)$, $g_{tt}=g_{tt}\left(r\right)$ and $g_{rr}=g_{rr}\left(r\right)$ are functions of radius $r$. 
As we will show below, the energy-momentum and metric tensors are both of isotropy. 

Then consider the transformation from the Cartesian coordinates $\{x^i\}$ to the polar coordinates $\{r,\theta,\varphi\}$:
\begin{eqnarray}
\label{eq:Pcoordinates}
\begin{array}{rcl}
\displaystyle
x=r\sin\theta \cos\varphi, \quad\,y=r\sin\theta \sin\varphi,~~~z=r\cos\theta.
\end{array}
\end{eqnarray}
This implies (see Weinberg's book~\cite{Weinberg:1972kfs} for details) 
\begin{eqnarray}
\label{eq:xidxi}
\begin{array}{rcl}
\displaystyle
x_{i}\,\mathrm{d} x^{i}\!&=&\!r\,\mathrm{d} r,\quad\mathrm{d} x_{i}\otimes\mathrm{d} x^{i}\!=\!\mathrm{d} r\otimes\mathrm{d} r\!+\!r^2 \mathrm{d} \theta\otimes\mathrm{d} \theta\!+\!r^2\sin^2\theta\,\mathrm{d} \varphi\otimes\mathrm{d} \varphi, 
\end{array}
\end{eqnarray}
with $\otimes$ being the tensor product, which are both rotational invariants. 
Think of isotropy as invariance under rotations, suitably generalized in general relativity~\cite{Carroll2014}.
Thus, the most general form of the static, isotropic tensor of rank two can be definitively expressed in terms of these two rotational invariants as bases. 

Combining equations~\eqref{eq:Tcomponets} and~\eqref{eq:sphcartesian}, we derive
\begin{eqnarray}
\label{eq:Tlk}
\begin{array}{rcl}
\displaystyle
T_{tt}=&+&g_{tt}\left(r\right)\,A(r),~~~
T_{ti}=0,\\[2mm]
T_{ij}=&-&\bigg[g_{rr}\left(r\right)C(r)+\left(\frac{g_{rr}\left(r\right)-1}{r^2}\right)B\left(r\right)\bigg]x_{i}\,x_{j}-\,B\left(r\right)\delta_{ij}, 
\end{array}
\end{eqnarray} 
by which we define the energy-momentum tensor~\eqref{eq:fullT} and find that the tensor does not depend on $t$, and depends on $x^i$ and $\mathrm{d} x^{i}$ only through the two rotational invariants in~\eqref{eq:xidxi}. 
Actually, we can further rewrite the energy-momentum tensor in the following form, 
\begin{eqnarray}
\label{eq:Tfull}
\nonumber
\widetilde{T}&=&g_{tt}\left(r\right)\,A(r)\,\mathrm{d} t\otimes\mathrm{d} t-g_{rr}\left(r\right)\bigg[B\left(r\right)+C(r)\,r^{2}\bigg]\mathrm{d} r\otimes\mathrm{d} r \\[2mm]
&-&\,B\left(r\right)\left(r^2 \mathrm{d} \theta\otimes\mathrm{d} \theta+r^2\sin^2\theta\,\mathrm{d} \varphi\otimes\mathrm{d} \varphi\right),
\end{eqnarray}
which is already the most general form in terms of the bases given by~\eqref{eq:xidxi} that $\widetilde{T}$ can take in the polar coordinates. 
Correspondingly, the metric~\eqref{eq:sphcartesian} can be expressed as 
\begin{eqnarray}
\label{eq:gsph}
\begin{array}{rcl}
\displaystyle
\mathrm{d} s^{2}\!=\!g_{tt}\left(r\right)\mathrm{d} t^{2}\!-\!g_{rr}\left(r\right)\mathrm{d} r^{2}\!-\!r^2 \left(\mathrm{d} \theta^2\!+\!\sin^2\theta\,\mathrm{d} \varphi^2\right),
\end{array}
\end{eqnarray}
where $t$ is the time-like coordinate while others are space-like coordinates. This metric is consistent with that shown by Kiselev in \cite{Kiselev:2002dx}. 
Thus, both the energy-momentum and metric tensors defined by equations~\eqref{eq:Tcomponets} and~\eqref{eq:sphcartesian} are spherically symmetric. 
Actually they have the same form as the general static isotropic tensor of rank two, 
which is referred to as the {\it standard form} of the isotropic tensor by Weinberg in \cite{Weinberg:1972kfs}.
Therefore, the energy-momentum tensor~\eqref{eq:Tcomponets} already takes the most general isotropic form in a static, isotropic spacetime.

\begin{figure}
\centerline{
\includegraphics[width=0.68\columnwidth]{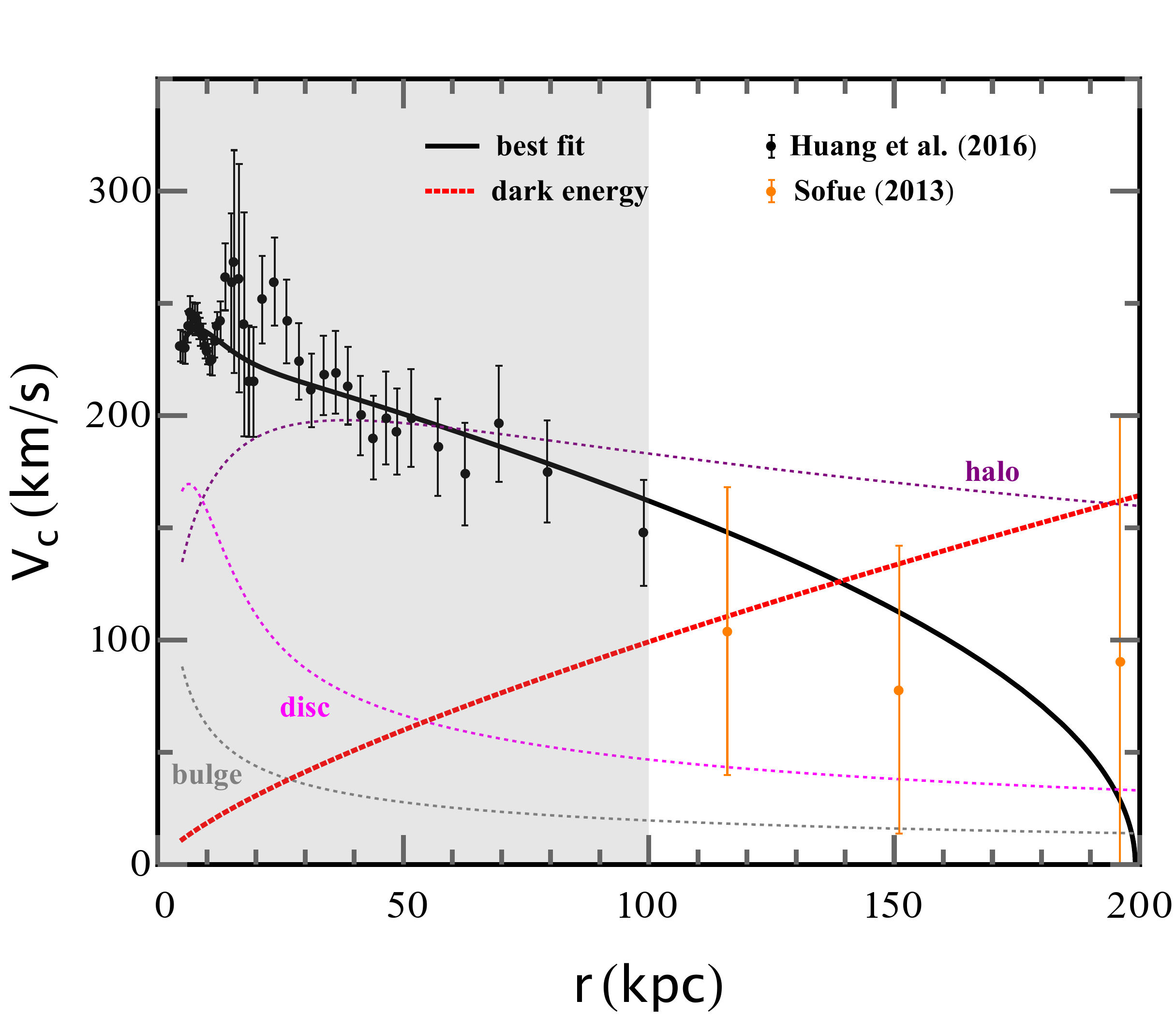}
}
\caption{Fitting to the MW rotation curve. 
Here we adopt data from~\cite{2016MNRAS.463.2623H} (black dots), and~\cite{Sofue:2013kja} (orange dots). 
The bulge, disc, halo, and DE components are shown as the dotted lines in grey, magenta, purple and red, respectively. 
The unshaded region represents the outer MW region with $r\gtrsim100~\mbox{kpc}$. In particular the red line stands for the negative contribution to the MW rotation curve from the dark force. 
}
\label{fig:component}
\end{figure}

In general relativity, a perfect fluid is defined by the condition that, at any spacetime point $P$, there is always a locally inertial frame, 
which is a sufficiently small region around $P$, comoving with the fluid element, 
and in which the fluid is isotropic for the comoving observer at $P$. 
More exactly, as seen from the comoving observer, 
the energy–momentum tensor of the fluid at the point $P$ should satisfy $T^{t}_{\,\,\,i}=0$ and $T^{i}_{\,\,\,j}\propto\,\delta^{i}_{\,\,\,j}$. 
Note that the tetrad of the comoving observer depends on the coordinate system chosen, and hence also the measured values by the same observer.
By this definition, the fluid described by~\eqref{eq:Tcomponets} can be treated to be perfect.
In fact, following the same steps as outlined in \cite{Weinberg:1972kfs}, the energy-momentum tensor~\eqref{eq:Tcomponets} can always be expressed in the special form of $T^{t}_{\,\,\,i}=0$ and $T^{i}_{\,\,\,j}\propto\,\delta^{i}_{\,\,\,j}$ under coordinate transformations. Thus, this fluid is isotropic for the comoving observer in some coordinate system. 
In the same coordinate system, the energy-momentum tensor~\eqref{eq:Tcomponets} can directly associate with a DE model\footnote{Nevertheless the two solutions cannot be combined directly through coordinate quantities. Usually, the coordinates used in one solution differs in the physical meaning from those used in another one, although the two systems of coordinates can be expressed using the same set of symbols. In general relativity, the two solutions can only be combined into a single one through their physical quantities, like the dynamical pressure~\eqref{eq:rho-pressure} and the EoS parameter~\eqref{eq:EOS}, rather than their coordinate quantities.} where $\widetilde{T}$ of the DE candidate takes the form ${\rm diag} (\rho, -p, -p, -p)$ in that coordinate system.

Generally, the specific form of the energy-momentum tensor changes from one coordinate system to another.
In classical fluid mechanics, the perfect fluid has the form of an identity matrix.
However, in general theory of relativity, the stress tensor of the perfect fluid can not always take that form, especially under general coordinate transformations.
In physics, the concept of a perfect fluid should be introduced independently of the coordinate system chosen.
To arrive at this concept, we need not restrict the stress tensor to the form of the identity matrix in a curved spacetime. 
Additionally, the energy-momentum tensor of the perfect fluid should be form-invariant under coordinate transformations,
Thus, it needs to take a more general form.
Furthermore, it should satisfy the isotropy condition, somewhat generalized in general relativity \cite{Rindler2006book,Carroll2014}. 
In fact, the energy-momentum tensor given by \eqref{eq:Tcomponets} has already taken the most general isotropic form and meets all these requirements. 
Therefore, the fluid with the energy-momentum tensor of the same form \eqref{eq:Tcomponets} can be defined as the $\textit{perfect fluid}$ in general relativity.

Now come back to DE in a real astrophysical system. Its EoS parameter $w$ can be treated as a constant.
If we impose the condition of $w={\rm constant}$, the energy-momentum tensor~\eqref{eq:Tcomponets} reduces to 
\begin{align}
\label{eq:Tcomponets1} 
T^{0}_{\,\,\,0}=\rho_{w}=A(r),~~~T^{i}_{\,\,\,j}=\lambda\,\rho_{w}\bigg[\widetilde{B}\,\delta^{i}_{\,j}-\left(1+3\,\widetilde{B}\right)\frac{x^{i}x_{j}}{r^2}\bigg],
\end{align}
where $\lambda$ is a constant parameter,
and $\widetilde{B}$ is an arbitrary parameter, depending on local properties of the DE candidate in the astrophysical system. 
As we will see, only if $\widetilde{T}$ is constructed in this way, we can keep the EoS parameter to be a constant. 
Thus, 
\begin{align}
\label{eq:Tcomponets2}
B\left(r\right)=\lambda\,\rho_{w}\,\widetilde{B},~~~C\left(r\right)=-\lambda\,\rho_{w}\bigg[1+3\,\widetilde{B}\bigg]\frac{1}{r^2},
\end{align}
which\footnote{ Note that $\widetilde{B}=-(1+3w)/6w$. If $w=-1$, one always has $C(r)=0$.
In this special case, as shown by equation~\eqref{eq:Tcomponets}, the off-diagonal components of the energy-momentum tensor vanish,
and then the energy-momentum tensor takes the diagonal form ${\rm diag} (\rho, -p, -p, -p)$.  
However, when $w\neq-1$, the non-zero off-diagonal components begin to appear. 
Thus, the energy-momentum tensor no longer take a diagonal form. Obviously, it cannot directly represent the solution for a cosmological DE model.
In this general case, the energy-momentum tensor can only be associated with a DE model by 
using the physical quantities defined in equations~\eqref{eq:rho-pressure} and~\eqref{eq:EOS}, as we have already shown. In fact, the association can only be made possible due to the definitions of these quantities, which has so-far been neglected.} takes the same form as that shown in the literature~\cite{Kiselev:2002dx}. 
By the definition~\eqref{eq:rho-pressure} and from equations~\eqref{eq:Tcomponets}, we obtain the energy density and dynamical pressure of the DE:
\begin{eqnarray}
\label{eq:edensity}
\rho_{w}&\equiv&A\left(r\right),\\
\label{eq:pdensity}
p_{w}&\equiv&-\frac{~\mathrm{Tr}\big[\,T^{i}_{\,\,\,j}\,\big]~}{3}=-\bigg[B\left(r\right)+\frac{1}{3}\,C\left(r\right)r^2\bigg],
\end{eqnarray}
which are defined in the same way as that in the ERW spacetime. 
Notice that the trace of all the components of any energy-momentum tensor is physical as it is independent of the choice of coordinates.
Physically, the energy density can be measured by the local observer.
In general, $\rho_{w}$ is fully determined by a specific DE model and it is unchanged in different coordinates.
Thus, the coordinate transformation is only on the components of the stress tensor.
Under such a transformation, the dynamical pressure is invariant as it is equivalent to the trace of the components of the stress tensor.

Substituting~\eqref{eq:Tcomponets2} into~\eqref{eq:edensity} and~\eqref{eq:pdensity}, one derive the equation of state 
\begin{eqnarray}
\label{eq:pdensityw}
p_{w}=\frac{\lambda}{3}\,\rho_{w}, 
\end{eqnarray}
which determines the DE's influence on what happens in an astrophysical system. 
Thus, we obtain the EoS parameter, i.e., $w=\lambda/3$, which is a constant parameter. 
Then, repeating the same derivations as in~\cite{Kiselev:2002dx,He:2017alg}, we can obtain the SdS$_{w}$ metric~\cite{Zhang:2021ygh} 
for the isolated system of a point-like object with mass $M$ from solving the Einstein equation,
\begin{eqnarray}
\label{eq:dSS2}
\begin{array}{rcl}
\displaystyle
\mathrm{d} S_{w}^2 \!=\!&+& \!\!\big[1\!-\!2\,\frac{\,M\,}{r}\!-\!2\left( \frac{\,r_{\!o}^{}\,}{r} \right)^{\!3w+1}\big]\mathrm{d} t^{2}-\!\!\frac{1}{\, \big[1\!-\!2\,\frac{\,M\,}{r}\!-\!2\left( \frac{\,r_{\!o}^{}\,}{r} \right)^{\!3w+1}\big] \,}\mathrm{d} r^2- r^{2}\left(\mathrm{d} \theta^2+\sin^{2}\theta\,\mathrm{d} \varphi^2\right),~~~
\end{array}
\end{eqnarray}
where $\,r_o^{}\,$ is a parameter that is fully determined by the chosen cosmological DE model and characterizes a specific cosmological scale at redshift $z_{r}$ in that model. 
Especially in the CC model, one can derive the value of $r_{\rm{o}}$ as $r_{\rm{o}}=\sqrt{6/\Lambda}$~\cite{Ho:2015nsa,He:2017alg}. 
Here, the geometrized unit system ($G=c=1$) is adopted throughout. As shown in~\cite{Zhang:2021ygh},
there is only one $w$-term responsible for the contributions of various forms of DE candidates.

The SdS$_{w}$ metric exhibits a standard form of the static isotropic tensor of rank two, which is identical to that presented by Weinberg in \cite{Weinberg:1972kfs}. 
This property indicates that the metric is isotropic in the framework of general relativity. 
In the special case of $M=0$, the SdS$_{w}$ metric is quasi-Minkowskian, as the $w$-term becomes negligible at small enough $r$, 
such as on galactic scales where it can be ignored to an accuracy of $\sim10^{-6}$.
Therefore, the isotropic energy-momentum tensor of the form \eqref{eq:Tcomponets} has been constructed based on a set of quasi-Minkowskian coordinates.
As previously stated, the isotropic energy-momentum tensor for a fluid generally implies that the fluid is intrinsically a perfect one.
Nevertheless, it should be noted that the isotropy of the energy-momentum tensor does not necessarily imply that the stress tensor can always take the form of an identity matrix.
For instance, in the SdS$_{w}$ case where $w\neq-1$, if imposing the condition $T^{r}_{\,\,\,r}=T^{\theta}_{\,\,\,\theta}=T^{\varphi}_{\,\,\,\varphi}$,  
we cannot find any solution to the Einstein equation.  
Conversely, in the same polar coordinate system, the Einstein equation dictates that $T^{i}_{\,\,\,i}$ varies with $i$.
This means that DE must, to some degree, redistribute itself in response to the gravitational influence of a mass distribution.
However, DE can still be described as a perfect fluid; that is, its energy-momentum tensor can still take the standard Weinberg's isotropic form.

Usually, the EoS parameter of DE cannot be defined in analogy to what we have done for the traditional perfect fluid in fluid mechanism.
Nevertheless, in the SdS$_{w}$ spacetime, the EoS parameter $w$ can be defined as the ratio of the dynamical pressure $p_{w}$ to the energy density $\rho_{w}$, regardless of the coordinate system adopted, using equations~\eqref{eq:rho-pressure} and~\eqref{eq:EOS}. Additionally, it is worth noting that $\rho_{w}$ and $p_{w}$ remain invariant even if the coordinate system is changed.
More exactly, once $\rho_{w}$ is known for a specific DE model, 
the stress tensor can be transformed into a form where $T^{i}_{\,\,\,j}\propto\delta^{i}_{\,\,\,j}$ under coordinate transformations, 
while the value of $p_{w}$ defined by equation~\eqref{eq:rho-pressure} remains unchanged. 
In the coordinate system that corresponds to the tensor form $T^{i}_{\,\,\,j}\propto\delta^{i}_{\,\,\,j}$, 
we can directly relate the SdS$_{w}$ metric with the solution for the DE model. 
Moreover, $\rho_{w}$ and $p_{w}$ are not affected by the specific form of the energy-momentum tensor and maintain their values.
However, it is improper to define an EoS parameter directly based on one of the components of the stress tensor in any coordinate system, even in which the stress tensor is already expressed in the form of an identity matrix\footnote{In general, the components of the stress tensor vary between different coordinate systems. Under some coordinate transformations, these components can be reexpressed to be different while the form of the stress tensor kept to be invariant.}, 
as this may bring the coordinate dependence into the definition of the EoS parameter.
Anyway, the EoS parameter is defined physically and globally in any curved, static spacetime via equation~\eqref{eq:EOS}, and it can be directly linked to the DE parameter $w$ in cosmology.

In summary, DE can be characterized by an evolving EoS parameter with the cosmological redshift, namely $w = w\left(z_{r}\right)$. 
The values of $w$ vary in different DE models, with each model describing DE using a specific $w$ value. For example, the CC model corresponds to $w=-1$, quintessence to $-1<w <-\frac{1}{3}$, and phantom to $w<-1$. On astrophysical scales, however, $w$ can be treated as a constant. Based on this assumption, the Einstein equation can be solved to obtain the SdS$_{w}$ metric, which can be used to describe different DE effects on astrophysical scales.


\subsection{Newtonian analogy}\label{NAnalogy}

If a static spacetime is spherically symmetric, its metric can be written in the following form,
\begin{eqnarray}
\begin{array}{rcl}
\displaystyle
\nonumber
\mathrm{d} s^2\!=\!\left(1\!+\!2\,\Phi\right)\mathrm{d} t^2\!-\!\left(1\!+\!2\,\Phi\right)^{-1}\mathrm{d} r^2\!-\!r^2\left(\mathrm{d} \theta^2\!+\!\sin^2\theta\mathrm{d} \varphi^2\right).
\end{array}
\end{eqnarray}
In the special SdS$_{w}$ case, one has
\begin{equation}\label{eq:potential}
\Phi=-\frac{M}{r}-\left(\frac{r_{\rm{o}}}{r}\right)^{3w+1}\,,
\end{equation}
where the first term is the Newtonian term with mass $M\,$, and the second term arises directly from the DE contribution. 
Note that $3w+1<0\,$, which is required by the accelerated cosmic expansion. 
As illustrated in this equation, the DE term dominates at large radii. 
However, this is obtained by solving the Einstein equation for the isolated system of a single point-like mass. 
Therefore, equation \eqref{eq:potential} is applicable only to an isolated astrophysical system. In reality, it will begin to fail to describe the gravitational system at too large radii due to tidal effects from any other astrophysical object.

In the general thoery of relativity, the Einstein equation can be reexpressed as
\begin{equation}
\label{eq:EinsteinEqs}
R^{\mu}_{\,\,\,\nu}=8\pi \,(T^{\mu}_{\,\,\,\nu}-\frac{1}{2}\delta^{\mu}_{\,\,\,\nu} \,T)\,, 
\end{equation}
which consists of two parts, namely 
\begin{equation}
\label{eq:Tmade}
T^{\mu}_{\,\,\,\nu}=T^\mu_{\,\,\,\nu,\rm{m}}+T^\mu_{\,\,\,\nu,w},
\end{equation}
where $T^\mu_{\,\,\,\nu,\rm{m}}\,$ and $T^\mu_{\,\,\,\nu,w}$ represent the matter and DE contributions, respectively.

For the matter part, we have 
\begin{equation}
T^\mu_{\,\,\,\nu,{\mathrm m}}=\left(\rho_{{\mathrm m}},-p_{\mathrm m},-p_{\mathrm m},-p_{\mathrm m}\right), 
\end{equation}
with $p_{\mathrm m}\ll\rho_{{\mathrm m}}$, where $\rho_{{\mathrm m}}$ and $p_{\mathrm m}$ are the mass density and pressure of matter, respectively. 
Generally, the matter part obeys the Poisson equation,
\begin{eqnarray}
\label{eq:MaPoissonEqs}
\nabla^2 \Phi_\mathrm{m}&=&4\pi\,\left(\rho_\mathrm{m}+3\,p_\mathrm{m}\right)\\ 
\label{eq:MaPoissonEqs1}
&\simeq&4\pi\,\rho_\mathrm{m},
\end{eqnarray}
by which~\eqref{eq:MaPoissonEqs1} we can derive the Newtonian potential,
\begin{equation}
\label{eq:NewPotetical0}
\Phi_m(\vec{r})=-\int\frac{\rho_\mathrm{m}(\vec{r^\prime})}{\mid \vec{r}-\vec{r^\prime}\mid}\mathrm{d} \vec{r^\prime}\,, 
\end{equation}
where the mass density $\rho_\mathrm{m}(\vec{r^\prime})$ distributes over a space region. 
In a realistic astrophysical system, the matter potential can be described well by $\rho_\mathrm{m}(\vec{r^\prime})$ in this way within Newtonian gravity.
For matter, its pressure can be ignored compared to its mass density;
that is, the EoS parameter is nearly zero.
The total mass of matter is $M=\int\rho_\mathrm{m}(\vec{r})\mathrm{d} \vec{r}$.
If the mass distribution is point-like or spherically symmetric, the potential reduces to
\begin{equation}
\quad \Phi_\mathrm{m}=-\dfrac{M}{r}\,,
\end{equation}
with its mass density $\rho_\mathrm{m}(\vec{r})$ distributed over the space region inside radius $r$, which is in accordance with the Newtonian term shown in equation\,\eqref{eq:potential}.

Now come back to the DE part. According to the time-time component of the Einstein equation\,\eqref{eq:EinsteinEqs}, 
\begin{equation}
\label{eq:00EinsteinEqs}
R^{0}_{\,\,\,0}=8\pi \left(T^{0}_{\,\,\,0}-\frac{1}{2}T^{\mu}_{\,\,\,\mu}\right)\,, 
\end{equation}
one finds
\begin{equation}
\partial_r\partial_r\Phi_w+\frac{2}{r}\partial_r\Phi_w=8\pi\left(\frac{1+3w}{2}\rho_w\right),
\end{equation}
which is equivalent to the following Poisson equation,
\begin{eqnarray}
\label{eq:DEPoissonEqs}
\nabla^2 \Phi_w&=&4\pi\,\left(\rho_w+3\,p_w\right)\,.
\end{eqnarray}
where $\rho_{w}$ and $p_{w}$ are presented in equations~\eqref{eq:edensity} and~\eqref{eq:pdensity}.
Note that this equation holds exactly in general relativity.

The total potential $\Phi$ can be divided into the matter and DE parts, namely 
\begin{eqnarray}
\Phi&\equiv&\Phi_\mathrm{m}+\Phi_w\\
&=&-\int\frac{\rho_\mathrm{m}(\vec{r^\prime})}{\mid \vec{r}-\vec{r^\prime}\mid}\mathrm{d} \vec{r^\prime}-\left(\frac{r_{\rm{o}}}{r}\right)^{3w+1},
\label{eq:totalPotential}
\end{eqnarray}
where each part takes a model-independent form.
Correspondingly, we have 
\begin{equation}
\rho=\rho_\mathrm{m}+\rho_{w},\,\quad\,p=p_\mathrm{m}+p_{w},
\end{equation}
where $\rho$ and $p$ are the total density and pressure of both matter and DE, respectively.
Accordingly, we can derive the Poisson equation,
\begin{equation}
\label{PoEtotal}
\nabla^2 \Phi=4\pi(\rho+3\,p)\,,
\end{equation}
which include both the matter and DE contributions.
Thus, according to the additivity and linearity of the Poisson equation, we derive the most general form of the total potential, $\Phi=\Phi(\vec{r})$, exactly as shown in equation~\eqref{eq:totalPotential}.
The total potential $\Phi$ can be applied directly to real situations, without any further restrictions.
As it shows, the matter and DE terms are on the same footing;
both of them obey the Poisson equation~\eqref{PoEtotal}. 
However, there are differences between the two terms. For instance,
the matter term induces an attractive force, whereas the DE term generates a repulsive force, referred to as the {\it dark force} or known as the {\it fifth force} in the literature~\cite{Zhang:2021ygh}.


\subsection{The repulsion of the dark force}\label{effect}

In the following, we will discuss the dark force and its effects on galactic scales. 
Usually, the typical galactic scale is about $\sim100$ kpc. 
For example, the MW has a virial radius of $r_{\mbox{\scriptsize vir}}\sim260~\mbox{kpc}$~\cite{2016MNRAS.463.2623H}, 
which is often used to characterize the size of the MW galaxy. Given that rotation velocities (RVs) of stars or gas clouds bound to the MW is around $200~\mbox{km/s}\ll c\,$, relativistic effects can be neglected. 
The total potential is very weak, i.e., $\Phi\ll\,1$, so the weak field approximation holds well. 
In the MW region with $r\lesssim\,r_{\mbox{\scriptsize vir}}$, this is indeed a good approximation. Exactly, one has
\begin{equation}\label{eq:weak}
\mid\!\Phi_w\!\mid\,\lesssim\,\mid\!\Phi_\mathrm{m}\!\mid\,\sim\frac{M}{r}\sim V^2\lesssim10^{-6}\ll1\,. 
\end{equation}
In fact, this is supported by the fact that the outer MW part is not torn apart by the dark force. 
Denote $a=a(z)$ as the cosmological expansion factor.
For the MW, we have $\frac{|\Delta a|}{a}\sim\frac{|\Delta z|}{1+z}\sim Hr_{\mbox{\scriptsize vir}}=6\times10^{-5}$.
Thus, the evolution of $w$ with $z_{r}$ is negligible; in other words, $w$ can be regarded as a constant on the MW scale. 
So we can neglect the cosmic expansion effect on $\Phi$. 
Additionally, under the weak field approximation, the potential $\Phi$ can be treated as a traditional gravitational potential in analogy to what we have done in Newtonian gravity. 
Therefore, the existence of DE modifies the specific form of the potential.

The DE contribution gives rise to a correction term $\Delta\Phi$ in the gravitational potential, namely $\Delta\Phi=\Phi_w$,
Now we consider its induced force. Generally, one has 
\begin{equation}\label{eq:force}
\vec{F}=-\vec{\nabla} \Phi=-\vec{\nabla} \Phi_\mathrm{m}-(3w+1)\,\frac{1}{r}\,\left(\frac{r_{\rm{o}}}{r}\right)^{3w+1}\hat{e}_r,
\end{equation}
which holds well in the weak field approximation. 
For a point-like mass $M$ or for regions outside a
spherically symmetric mass-distribution,
the first term in equation\,\eqref{eq:force} is closely related to the total mass $M=M(r)$ within a radius of $r$ from the mass center by
\begin{equation}
-\vec{\nabla} \Phi_\mathrm{m}=-\frac{M}{r^2}\,\hat{e}_r\,,
\end{equation}
which is just the attractive Newtonian force.
The second term in equation~\eqref{eq:force} comes directly from the DE contribution.
The repulsive dark force can be generated by this term.
In the weak field approximation, the dark force shown by~\cite{Zhang:2021ygh} takes the same form as the second term in equation~\eqref{eq:force}.
Here, the dark force takes a model-independent form;
exactly, it can well describe various DE models characterized by different $w$ values: $w=-1$ for the CC model, $-1<w<-\frac{1}{3}$ for the quintessence model, and $w<-1$ for the phantom model.

For an isolated astrophysical system, the appearance of the dark force is unavoidable, and the repulsion must be included into the total gravitational force, 
together with the matter contribution.
Interestingly, the dark force has a negative contribution to the RV values, namely 
\begin{equation}
\Delta_{w} V^2(r)\equiv-\mid3w+1\mid\left(\frac{r_{\rm{o}}}{r}\right)^{3w+1}\,,
\end{equation}
where the DE contribution changes significantly with $w$. 
We hereby can investigate the contribution of the dark force to various rotation curves on astrophysical scales.

Requiring the cancellation between the two forces, one can derive the critical radius. Letting $M_{\mbox{\scriptsize cri}}=M(r_{\mbox{\scriptsize cri}})$, one gets
\begin{equation}
\quad r_{\mbox{\scriptsize cri}}= r_{\rm{o}}\left(\mid3w+1\mid\frac{r_{\rm{o}}}{M_{\mbox{\scriptsize cri}}}\right)^{\frac{1}{3w}}\,. 
\end{equation}
which exactly coincides with \cite{Zhang:2021ygh}.
In the special CC case, it reduces to
\begin{equation}
r_{\mbox{\scriptsize cri}}\,\Big{|}_{\Lambda}=\left(\frac{3GM_{\mbox{\scriptsize cri}}}{\Lambda}\right)^{\frac{1}{3}}\,,
\end{equation}
which agrees with~\cite{Ho:2015nsa}. The critical radius $r_{\mbox{\scriptsize cri}}$ is the typical scale of the dark force; usually, it decreases with $\mid\!w\!\mid$.
For the MW galaxy, $r_{\mbox{\scriptsize cri}}$ is about $\sim500~\mbox{kpc}$, which is estimated by using the CC model~\cite{Ho:2015nsa}.

For a point-like mass distribution, the dark force effect on any rotation curve can be characterized by
\begin{equation}\label{eq:scale}
\frac{\mid\Delta_{w} V^2(r)\mid}{V_{\mbox{\scriptsize N}}^2}=\left(\frac{r}{r_{\mbox{\scriptsize cri}}}\right)^{-3w}\,,
\end{equation}
where $V_{\mbox{\scriptsize N}}=V_{\mbox{\scriptsize N}}(r)$ is the RV value obtained from the purely Newtonian force. 
The DE correction to the RV value is $\frac{\mid\Delta_{w} V(r)\mid}{V_{\mbox{\scriptsize N}}}=\frac{1}{2}\frac{\mid\Delta_{w} V^2(r)\mid}{V_{\mbox{\scriptsize N}}^2}$.
Setting $w=-1\,$, one gets $\frac{\mid\Delta_{w} V(r)\mid}{V_{\mbox{\scriptsize N}}}\ge3\%$ at $r=0.4\,r_{\mbox{\scriptsize cri}}\,$. 
Here we assume that all the matter distributes in the region within $r=0.4\,r_{\mbox{\scriptsize cri}}$.
This ignores the mass distribution in the outer region with $r\ge0.4\,r_{\mbox{\scriptsize cri}}$.
Using equation~\eqref{eq:scale}, we therefore set a lower limit on the matter contribution to the rotation curve.

For a spherically symmetric mass distribution $M(r)$, if assuming $M(r)\propto r$ in the outer region with $r\ge0.4\,r_{\mbox{\scriptsize cri}}$, 
we find that $V_{\mbox{\scriptsize N}}$ does not change with $r$.
In fact, the rotation curve in the outer region is not as flat as we shown under that assumption. 
Usually it deceases slightly with $r$, as illustrated in Figure~\ref{sample}. 
This means that we set an upper limit on the matter contribution to the rotation curve. 
In this regime, the dark force contributes to the rotation curve at the least level. Exactly, one has
\begin{equation}\label{eq:scale2}
\frac{\mid\Delta_{w} V^2(r)\mid}{V_{\mbox{\scriptsize N}}^2}
=\left(\frac{r}{r_{\mbox{\scriptsize cri}}}\right)^{-3w-1},
\end{equation}
which suggests reliance of the DE correction on $r_{\mbox{\scriptsize cri}}$ rather than on $r_{\mbox{o}}$.
Within the critical radius $r_{\mbox{\scriptsize cri}}$, the effect of the dark force on the rotation curve gets enhanced significantly as $w$ increases. 
According to the CC model, the DE correction $\frac{\mid\Delta_{w} V(r)\mid}{V_{\mbox{\scriptsize N}}}$ is about 
$\sim2\%$ for $r=0.2\,r_{\mbox{\scriptsize cri}}$ and about $\sim8\%$ for $r=0.4\,r_{\mbox{\scriptsize cri}}$. 
Note that the present precision of the MW rotation curve is about $\sim3\%-8\%$ within $r=0.2\,r_{\mbox{\scriptsize cri}}\sim100~\mbox{kpc}$~\cite{2016MNRAS.463.2623H}.
Therefore, the effects of the dark force on the rotation curve are large enough to be detected at this precision level.

Once going beyond the critical radius $r_{\mbox{\scriptsize cri}}$, any objects from an astrophysical system cannot be treated as being gravitationally bounded to the system.
Usually, those objects will be influenced significantly by nearby systems.
In a real-world scenario, the gravitational effects of neighboring systems can become non-ignorable at $r\lesssim\!\,r_{\mbox{\scriptsize cri}}$ \cite{Ho:2015nsa}.
Therefore, we need to introduce an effective radius, denoted by $r_{\mbox{\scriptsize eff}}$, 
to define an ideal isolated region where the gravitational effects from any other systems can be ignored.
Let $n_{\mbox{\scriptsize eff}}=r_{\mbox{\scriptsize eff}}/r_{\mbox{\scriptsize cri}}$. In general, it ranges from $\sim0.2$ to $\sim1.7$~\cite{He:2017alg}.
For an ideal isolated system, the upper limit of $n_{\mbox{\scriptsize eff}}$ is $\sim1.7$ in the CC model~\cite{He:2017alg}. 
The index $n_{\mbox{\scriptsize eff}}$ can be used to judge the extent to which the chosen region is gravitationally isolated from surrounding regions or systems.
Its value can be chosen properly to avoid gravitational effects from nearby astrophysical systems, so that the region with $r\lesssim\!\,r_{\mbox{\scriptsize eff}}$ behaves as an isolated system.


\section{Data and fits}
\label{Danalysis}

To obtain the rotation curves in the MW galaxy, we use two sets of data: one for the inner MW region with $r\sim4.5-100~\mbox{kpc}$ and the other one for the outer MW region with $r\sim100-200~\mbox{kpc}$. The data for $r<100~\mbox{kpc}$ come from~\cite{2016MNRAS.463.2623H}, 
while for $r>100~\mbox{kpc}$ from~\cite{1992MNRAS.255..105K,Sawa:2004tx,Sofue:2013kja}. 
At larger radii than $\sim200~\mbox{kpc}$, there seems to be an increase in the RV value. 
Especially, beyond $r\sim r_{\mbox{\scriptsize vir}}$,  the rotation curve shows an abnormal rise, 
indicating that the influence of nearby galaxies become strong.
This is also the reason why we often choose the viral radius to characterize the outer MW boundary.
To minimize the influence of nearby galaxies, we choose $n_{\mbox{\scriptsize eff}}=0.4$ to define an isolated region. 
Accordingly, we only use the data points between $4.5-200~\mbox{kpc}$ to carry out a data analysis.

For constructing the rotation curve in the outer MW region, beyond the Galactic disc, we have to rely on non-disc tracers like satellite galaxies that do not exhibit systematic motion~\cite{1992MNRAS.255..105K,Sawa:2004tx}.
For instance, it is the dwarf galaxies of the MW that allow us to extend the rotation curve to $\sim 200$ kpc~\cite{2012AJ....144....4M}. 
These non-disc tracers move in various non-circular orbits around the MW center. This may yield systematic uncertainties in the RV measurements. 
However, it may be sufficient to describe the galactic structure for a first approximation~\cite{Sofue:2013kja}. 
Further dynamical studies on non-disc tracers will help us to understand these orbits and develop effective approaches 
for obtaining accurate RVs with reliable uncertainties at large radii~\cite{Jenkins:2020blc,Oakes:2022xcv}.
In addition we can also start long pointed observations of the non-disc tracers like satellite galaxies using current or future telescopes,
and make precise measurements for these non-circular orbits.
Anyway, to illustrate how to measure $w$ through rotation curves, we only use the published rotation curve data.

\subsection{Benchmark}\label{benchmark}
As shown by the second term in equation~\eqref{eq:force}, the dark force takes a model-independent form 
\begin{equation}
\vec{F}_w=-\vec{\nabla} \Phi_w=-(3w+1)\frac{r_{\rm{o}}^{3w+1}}{r^{3w+2}}\hat{e}_r\,,
\end{equation}
where the parameter $r_{\rm{o}}$ may vary in the different models~\cite{Zhang:2021ygh}.
In the CC model, one has $\Lambda=4.24\times10^{-66}~\mbox{eV}^2$~\cite{Aghanim:2018eyx}.
Thus, $r_{\rm{o}}=\sqrt{\frac{6}{\Lambda}}=7.71\times10^6~\mbox{kpc}$~\cite{Ho:2015nsa,Zhang:2021ygh,He:2017alg},
which is taken as a benchmark point in this work.

Before a specific data fit, 
there is no reason to believe that we already know the mass distribution in the outer MW region very well.
Thus, we cannot estimate the critical radius directly if there is not any assumption made for the dark force. 
However, we can make an order-of-magnitude estimate of the critical radius. 
To do this, we need to assume that in the outer MW region, the Newtonian RV value $V_{\mbox{\scriptsize N}}=V_{\mbox{\scriptsize N}}(r)$ keeps to be of the same order of magnitude 
as that at $r=r_{0}$; that is, $V_{\mbox{\scriptsize N}}(r)\sim\,V_{0}=V_{\textrm{N}}(r_{0})$,
where $V_{\mbox{\scriptsize N}}$ is the RV value derived from the purely Newtonian force.
In this case, by definition, the critical radius can be roughly estimated as
\begin{equation}\label{eq:critical radius}
V_{0}^2\sim\left|3w+1\right|\left(\frac{r_{\rm{o}}}{r_{\mbox{\scriptsize cri}}}\right)^{3w+1}\,. 
\end{equation}
For the MW galaxy, we set $r_{0}=80~\mbox{kpc}$. As shown in Figure~\ref{sample}, $V_{0}\approx180~\mbox{km/s}$.
In the region with $40~\mbox{kpc}\lesssim r \lesssim 80 ~\mbox{kpc}$, the Newtonian RV value $V_{\mbox{\scriptsize N}}$ is dominated by the contribution of the MW halo,
resulting in a plateau in the rotation curve over this region. Accordingly, we have
\begin{align}
V^2(r)\sim  V_{0}^2-\left|3w+1\right|\left(\frac{r_{\mbox{o}}}{r}\right)^{3w+1}\sim V_{0}^2\left[1-\Big(\frac{r_{\mbox{\scriptsize cri}}}{r}\Big)^{3w+1}\right]\approx V_{0}^2\,,
\end{align}
where the $w-$dependent term can be ignored in the region. 
Additionally, the Newtonian RV value $V_{\mbox{\scriptsize N}}$ shows a variation of less than $20\%$ in the region with $80 ~\mbox{kpc}\lesssim r \lesssim 260~\mbox{kpc}$ when the MW halo is estimated using the NFW model. Therefore, equation \eqref{eq:critical radius} holds very well in the range of $r\sim40-260~\mbox{kpc}$. 
For the MW galaxy, it is reasonable to set $V_{0}=180~\mbox{km/s}$. Then, the critical radius is about $\sim3300~\mbox{kpc}$ for $w=-1.0$, and $\sim150~\mbox{kpc}$ for $w=-0.8$. 
This means that if $w$ is as small as $-1.0$, the dark force will have little influence on the MW galaxy, 
whereas if $w$ is as large as $-0.8$, the outer MW part with $r\gtrsim150~\mbox{kpc}$ would be torn apart by the dark force before the MW galaxy could form. 
So $w$ cannot be too large, namely $w\lesssim-0.8$; otherwise, the matter in the outer region would overcome the net force and escape from the MW galaxy directly.

From equation~\eqref{eq:critical radius}, we can also estimate the ratio of the critical radius $r_{\mbox{\scriptsize cri}}$ to the cosmological scale $r_{\rm{o}}$ as
\begin{equation}\label{eq:V0Rcri}
\frac{r_{\mbox{\scriptsize cri}}}{r_{\rm{o}}}\sim\Bigg(\frac{V^2_0}{|3w+1|}\Bigg)^{\frac{1}{|3w+1|}}\equiv\frac{r_{\rm{\scriptsize df}}}{r_{\rm{o}}},
\end{equation}
where $r_{\rm{\scriptsize df}}$ is nearly equal to $r_{\mbox{\scriptsize cri}}$.
For the MW galaxy, $V_{0}/c\sim10^{-3}$. 
When choosing $-1.0\lesssim w\lesssim-0.8$, we find $10^{-5}\lesssim r_{\mbox{\scriptsize cri}}\,/\,r_{\rm{o}}\lesssim10^{-4}$. 
Thus, the value of $r_{\mbox{\scriptsize cri}}$ is mainly determined by the measured $V_{0}$. 
Conversely, we can use equation~\eqref{eq:V0Rcri} to estimate the $r_{\rm{o}}$ value cosmology-independently. 
The virial radius of the MW galaxy is $r_{\mbox{\scriptsize vir}}\sim260~\mbox{kpc}$~\cite{2016MNRAS.463.2623H}. As a result, $r_{\mbox{\scriptsize cri}}\gtrsim260$ kpc. 
Thus, it can expected that $r_{\mbox{\scriptsize cri}}\sim 260-1000~\mbox{kpc}$. 
Then we obtain $r_{\rm{o}}\sim\left(\frac{1}{3}-10\right)\,\sqrt{6/\Lambda}\sim2.6\times10^{6}-10^{8}~\mbox{kpc}$. 
So it is reasonable to set $r_{\rm{o}}\sim7.71\times10^6~\mbox{kpc}$ as a benchmark point.

As equation~\eqref{eq:V0Rcri} shows, the distance scale $r_{\rm{\scriptsize df}}$ can be used to approximate the critical radius through the observed $V_{0}$ value 
once a $w$ value is provided. With this equation, the dark force can be therefore rewritten as 
\begin{equation}\label{eq:FwV0rdf}
\vec{F}_w=\frac{V_{0}^{2}}{r}\,\left(\frac{r_{\rm{\scriptsize df}}}{r}\right)^{3w+1}\hat{e}_{r},
\end{equation}
where $V_0$ can be estimated from a galactic rotation curve.
It clearly indicates that the dark force strengthens significantly at the distance scale of $\sim r_{\rm{\scriptsize df}}$ from the galactic center, 
which is consistent with what is suggested by equation~\eqref{eq:scale}. 
In addition, we observe that $w$ represents how rapidly the strength of the dark force increases with radius.

\subsection{The galactic mass model}

Following~\cite{Calcino:2018mwh}, we assume that the MW mass distribution consists of three components, i. e., a bulge, a disc and a dark matter halo, and it extends to $\sim100-200~\mbox{kpc}$ continuously. 
Note that there are two localized dips at $r\sim11~\textrm{and}~19$ kpc in the MW rotation curve, respectively. 
Extra components could be included to interpret these small-scale structures in the innermost WM region with $r\lesssim20$ kpc.
However, our aim is to detect the dark force by the rotation curve fitting.
To the current precision of $\sim3\%-8\%$, the dark force begins to be detectable at $r\sim100\,\textrm{kpc}$,
which is much larger than the size of each localized dip or that of the innermost region. 
Therefore, the dark force is insensitive to these structures in the innermost region. 
So we can ignore their influence on the fitted values of $w$ and then parametrize the rotation curve as the three-component mass model with the bulge, disc and dark matter halo for the rotation curve fitting.
Then the Newtonian RVs contributed these three components can be given by
\begin{equation}
V_{\mbox{\scriptsize N}}^2=V_{\mbox{\scriptsize b}}^2+V_{\mbox{\scriptsize d}}^2+V_{\mbox{\scriptsize h}}^2\,. 
\end{equation}

There are three main MW components, which are described briefly below:

(a)~\textrm{The bulge}.
As shown in~\cite{McMillan:2011wd}, the bulge is close to axisymmetric. Note that its scale radius is about $r_{\mbox{\scriptsize cut}}\sim2.1~\mbox{kpc}$, 
which is much smaller than the characteristic radius that we are interested in, i.e., about $\sim100\,\textrm{kpc}$ . 
And as the radius increases, the internal structure of the bulge becomes quite unimportant. So we can approximately treat the bulge as a point-like mass of $M_{\mbox{{\scriptsize bulge}}}=8.9\times 10^9M_{\odot}\,$. Then,
\begin{equation}\label{eq:bulge}
\frac{V_{\mbox{\scriptsize b}}(r)}{\mbox{km/s}}=196\times\left(\frac{r}{\mbox{kpc}}\right)^{{-1/2}}\,. 
\end{equation}
In this way, $V_{\mbox{\scriptsize b}}=V_{\mbox{\scriptsize b}}(r)$ may be overestimated a bit. 
However, it cannot lead to a deviation $\Delta\,V_{\mbox{\scriptsize b}}$ of more than $\sim2\%$ from the estimated value of $V_{\mbox{\scriptsize b}}$ from the density profile presented in \cite{2016MNRAS.463.2623H}.
Namely, we have $\frac{\Delta\,V_{\mbox{\scriptsize b}}}{V_{\mbox{\scriptsize b}}}\lesssim2\%$ in the outer MW region with $r\geq100\,\textrm{kpc}$, 
which can be verified by strict numerical calculations. Given that $V_{\mbox{\scriptsize b}}$ is about one order of magnitude smaller than $V_{\mbox{\scriptsize N}}$ in this region, we can safely draw the conclusion that the deviation from the total RV value is less than $0.1\%$. Thus, for the sake of simplicity, we choose to characterize the bulge with a point-like mass model.

(b)~~\textrm{The disc}.
The disc component can be described by the surface-density profile~\cite{Dehnen:1996fa}
\begin{equation}
\Sigma_{\mbox{\scriptsize d}}(r)=\Sigma_{\mbox{\scriptsize d,0}}\exp(-r/r_{\mbox{\scriptsize d}})\,, 
\end{equation}
with a central surface density $\Sigma_{\mbox{\scriptsize d,0}}$ and a scale length $r_{\mbox{\scriptsize d}}\,$. 
Following~\cite{2016MNRAS.463.2623H}, we fix the local surface density of disc to be $54.4~M_{\odot}\,\mbox{pc}^{-2}$ at $r=8.34~\mbox{kpc}$ so as to match with the observations.
Then the circular velocity is given by
\begin{equation}\label{eq:disc}
V_{\mbox{\small c}}^ {2}(r) = 4\pi G\Sigma_{\mbox{\scriptsize d,0}}r_{\mbox{\scriptsize d}}y^{2}\Big[I_{0}(y)K_{0}(y)-I_{1}(y)K_1(y)\Big]\,,
\end{equation}
with $y=r/(2r_{\mbox{\scriptsize d}})\,$, where $I_n$ and $K_n$ ($n=0, 1$) are modified Bessel functions of the first and second kind, respectively.

(c)~~\textrm{The dark matter halo}.
We adopt the NFW density profile~\cite{Navarro:1996gj} to describe the dark matter halo: 
\begin{equation}\label{eq:halo}
\rho_{\mbox{\scriptsize h}}(r)=\rho_{\mbox{\scriptsize h,0}}(r/r_{\mbox{\scriptsize h}})^{-1}(1+r/r_{\mbox{\scriptsize h}})^{-2}\,, 
\end{equation}
Thus, the contribution to the circular velocity of the halo can be computed using
\begin{equation}
V_{\mbox{\scriptsize h}}^2=\frac{4\pi\rho_{\mbox{\scriptsize h,0}}r_{\mbox{\scriptsize h}}^3}{r}\left(\ln\frac{r_{\mbox{\scriptsize h}}+r}{r_{\mbox{\scriptsize h}}}-\frac{r}{r+r_{\mbox{\scriptsize h}}}\right)\,. 
\end{equation}
The actual profile of the dark matter density may differ from the NFW profile, but the asymptotic behavior of $r^{-3}$ is widely accepted~\cite{Ho:2015nsa}. 
Indeed, our results below will be insensitive to the density profile at small $r$.

\subsection{Circular velocities}
Finally, we also need to add to the circular velocity the DE correction term. Consequently, the total circular velocity $V_{\mbox{\small c}}$ can be rewritten as
\begin{equation}
V_{\mbox{\small c}}^2=V_{\mbox{\scriptsize N}}^2+\Delta_{w}V^2\,. 
\end{equation}

Up to now, we have parametrized the MW rotation curve.
In total, there are four free parameters used for fitting: one for the disc ($r_{\mbox{\scriptsize d}}$), two for the NFW halo ($\rho_{\mbox{\scriptsize h,0}}\,$, $r_{\mbox{\scriptsize h}}$), and one for the dark force ($w$). 
In the fitting procedure, we use the Levenberg-Marquardt algorithm to find the best-fitting values of the four parameters. 
Furthermore, a Markov chain Monte Carlo technique is used to sample the likelihood of the data~\cite{2010CAMCS...5...65G}.
The likelihood function is defined as
\begin{equation}
\mathcal{L}=\prod\limits_{i=1}^{N}\frac{1}{\sqrt{2\pi}\sigma_{V_{\mbox{\small c}, r_i}^{\mbox{\scriptsize obs}}}}\exp\frac{-[V_{\mbox{\small c}, r_i}^{\mbox{\scriptsize obs}}-V_{\mbox{\small c}, r_i}^{\mbox{\scriptsize model}}(\hat{\theta})]^2}{2\sigma_{V_{\mbox{\small c}, r_i}^{\mbox{\scriptsize obs}}}^2}\,, 
\end{equation}
where $N$ is the number of data points used in our fit, $\sigma_{V_{\mbox{\small c}, r_i}^{\mbox{\scriptsize obs}}}$ is the uncertainty of the observed RV value $V_{\mbox{\small c}, r_i}^{\mbox{\scriptsize obs}}$, and $\hat{\theta}$ represents the four fitting parameters that we want to determine. Actually, we make use of the PYTHON package LMFIT~\cite{Foreman-Mackey:2012any} to find the best-fit values and compute the confidence intervals as well as estimate the upper and lower bounds on the rotation curve.

\begin{table}
\centerline{
\begin{tabular}{|c|cccc|c|}
\hline
Range&$r_{\rm{\scriptsize d}} {\rm(kpc)}$&$\rho_{\mbox{\scriptsize h,0}}$ ($M_{\odot}\,{\rm pc}^{-3}$)&$r_{\mbox{\scriptsize h}} {\rm(kpc)}$&$w$&$\chi^2_{\mbox{\scriptsize red}}$\\
\hline
4.5-200~kpc&$2.9_{-0.1}^{+0.2}$&$0.011_{-0.003}^{+0.002}$&$18_{-3}^{+1}$&$-0.82_{-0.01}^{+0.01}$&$0.85$\\
\hline
4.5-100~kpc&$2.8_{-0.1}^{+0.1}$&$0.006_{-0.002}^{+0.004}$&$24_{-6}^{+8}$&$-0.79_{-0.02}^{+0.01}$&$0.85$\\
\hline
\end{tabular}
}
\caption{Best-fit parameters of the disc, halo, and dark energy components, obtained from the rotation curves from $r\sim4.5$ to $200~\mbox{kpc}$ and to $100~\mbox{kpc}$, respectively. 
The errors are at 1$\sigma$, and the reduced $\chi^2\,$ are shown in the last column. }
\label{tab:fit}
\end{table}


\section{Results and discussion}\label{result}

\begin{figure}
\centerline{
\includegraphics[width=0.68\columnwidth]{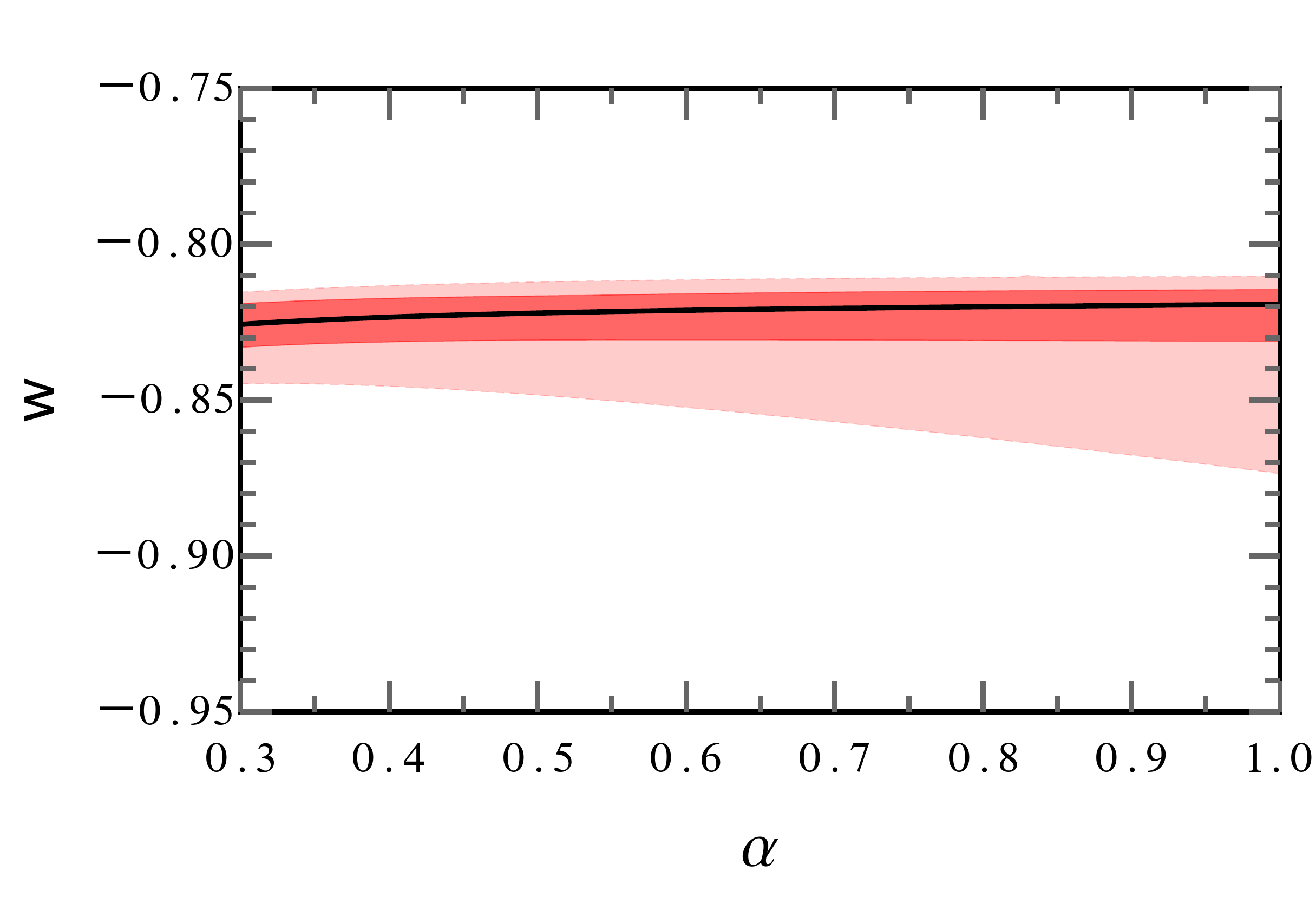}
}
\caption{The best-fit results of $w$ at different $\alpha$ values, where the uncertainties at $r\sim100-200~\mbox{kpc}$ are assumed to be $\alpha$ times the observed ones, with $\alpha$ ranging from 0.3 to 1. The black line represents the best-fit line. The 1$\sigma$ and 2$\sigma$ intervals are shown in dark and light red, respectively. }
\label{fig:NFW2sigma}
\end{figure}

By fitting the Huang et al.~2016 and Sofue~2013 data, we can obtain the parameters of the galactic mass model as well as the EoS parameter $w$.
The best-fit values of these parameters and their uncertainties are listed in Table~\ref{tab:fit}. 
We also illustrate in Figure~\ref{fig:component} the contributions from different MW components. 

For the disc, we adopt the same disc model as in~\cite{Calcino:2018mwh}. 
As expected, the fitted value of the disc parameter $r_{\mbox{\scriptsize d}}$ in this model agrees with that shown in~\cite{Calcino:2018mwh} within 1$\sigma$ errors.
For the dark matter halo, the best-fit halo parameters are determined to be $\rho_{\mbox{\scriptsize h,0}}=0.015_{-0.004}^{+0.007}~M_{\odot}\,\mbox{pc}^{-3}$ and $r_{\mbox{\scriptsize h}}=15_{-3}^{+4}$ kpc.
Note that the NFW model~\eqref{eq:halo} is widely used in previous works~\cite{2016MNRAS.463.2623H,Calcino:2018mwh}.
Within errors, our values of the halo parameters are also consistent with~\cite{2016MNRAS.463.2623H} and~\cite{Calcino:2018mwh}.

Previously, the local dark matter density $\rho_{\odot,\,\mbox{\scriptsize dm}}$ was determined at the Sun’s galactocentric radius $r=8.34~\mbox{kpc}$, 
without taking into account the influence of the dark force. 
Actually, it had been shown that $\rho_{\odot,\,\mbox{\scriptsize dm}}=0.32\pm0.02~\mbox{GeV cm}^{-3}$ in~\cite{2016MNRAS.463.2623H} 
and $\rho_{\odot,\,\mbox{\scriptsize dm}}=0.24^{+0.10}_{-0.09}~\mbox{GeV cm}^{-3}$ in~\cite{Calcino:2018mwh}, respectively.
On the other hand, from the newly fitted values of $\rho_{\mbox{\scriptsize h,0}}$ and $r_{\mbox{\scriptsize h}}$ in this work, 
we are able to obtain the local dark matter density $\rho_{\odot,\,\mbox{\scriptsize dm}}$ via equation\,\eqref{eq:halo}. 
Exactly, we have $\rho_{\odot,\,\mbox{\scriptsize dm}}=0.40_{-0.15}^{+0.07}~\mbox{GeV cm}^{-3}$, which coincides with the previous results within 1$\sigma$ errors.

The MW data between $r=4.5$ and $200~\mbox{kpc}$ provides a stringent constraint on the DE parameter $w$. 
Specifically, as shown in Table~\ref{tab:fit}, $w=-0.82_{-0.01}^{+0.01}$, which is obtained independently of any specific DE models.
If the EoS parameter $w$ is greater than $-0.85$, the strength of the repulsive dark force increases much more significantly with $w$ than the attractive Newtonian force does with the other parameters. In this particular case, the fitting results are more sensitive to $w$ than the other parameters. 
Therefore, compared to other parameters, there is a much tighter constraint on the DE parameter $w$.

Now there have been cosmological measurements on the EoS parameter for the dynamical DE candidate within $w(z_{r})$CDM, 
i.e., $w=-1.07_{-0.20}^{+0.21}$ at $z_{r}=0$~\cite{Zhao:2017cud},
where the evolution history of the EoS parameter, namely $w=w(z_{r})$, is reconstructed from a collection of cosmological data
by using a non-parametric Bayesian method based on applying a correlated prior~\cite{2012PhRvL.109q1301Z}. 
At 1$\sigma$ confidence level, it shows a little disagreement with our fitting value.
However, the two values coincide with each other within $\sim1.2\sigma$ errors. 
By measuring the effects of the dark force in the MW galaxy, we are able to cross-check and validate cosmological measurements to a certain extent.
Essentially, we have proposed a novel method to measure the DE parameter $w$, without relying on cosmological observations.

The dark force could have various effects on galaxies, and become strong at large galactocentric distances. 
As equation~\eqref{eq:scale2} shows, its effects on rotation curves increases with $r$ in the form of $\,r^{-3w-1}$. 
By choosing $r$ properly, these effects can be enhanced by several times within the effective radius $r=r_{{\rm eff}}$. 
However, the RV measurements at large distances are usually less precise than those at small distances. 
Denote $\Delta V_{\mbox{\scriptsize exp}}=\Delta V_{\mbox{\scriptsize exp}}(r)$ as the uncertainty of RVs at radius $r$.
As shown in Figure~\ref{fig:component}, the uncertainty $\Delta V_{\mbox{\scriptsize exp}}$ is $\sim20~\mbox{km/s}$ at $r\sim20-100~\mbox{kpc}$,  
and $\sim60~\mbox{km/s}$ at $r\sim100-200~\mbox{kpc}$. 
At present, the large uncertainties of the RV values at galactocentric distances pose challenges in detecting the dark force at a high level of precision.

More measurements will be done with current telescopes, like James Webb Space Telescope~\cite{Boccaletti_2015} and Very Large Telescope Interferometer~\cite{2017A&A...602A..94G}, as well as by future telescopes, such as China Space Station Telescope~\cite{2023arXiv230402196F}, European Extremely Large Telescope~\cite{2021JATIS...7c5005R}, Thirty Meter Telescope~\cite{2020SPIE11447E..2ZR}, Giant Magellan Telescope~\cite{2016SPIE.9908E..1UJ}, and Legacy Survey of Space and Time~\cite{LSST:2008ijt}, in the outer MW region with $r\sim100-200$~kpc. 
It can be expected that better and more data can be obtained by these telescopes, and the accuracy of measurements will be improved significantly in the near future. 

Assume the expected accuracy to be improved by a factor $\alpha$, namely $\Delta V_{\mbox{\scriptsize th}}(r)=\alpha\,\Delta V_{\mbox{\scriptsize exp}}$. 
Let us change the value of $\alpha$ and show how the dark force exerts its influence on the MW rotation curve to different accuracies.
If the factor $\alpha$ decreases to 0.3 from 1, $\Delta V_{\mbox{\scriptsize th}}$ at radius $r\sim100-200$~kpc will become compared with the observed uncertainties at $r\lesssim 100~\mbox{kpc}$.
Indeed, for each $\alpha$ value, we fit the MW rotation curve and obtain the fitting result.
As shown in Figure~\ref{fig:NFW2sigma}, the best-fit value of $w$ remains relatively constant with changes in $\alpha$, whereas the uncertainty in the $w$ value decreases significantly as $\alpha$ decreases.

\begin{figure}
\centerline{
\includegraphics[width=0.68\columnwidth]{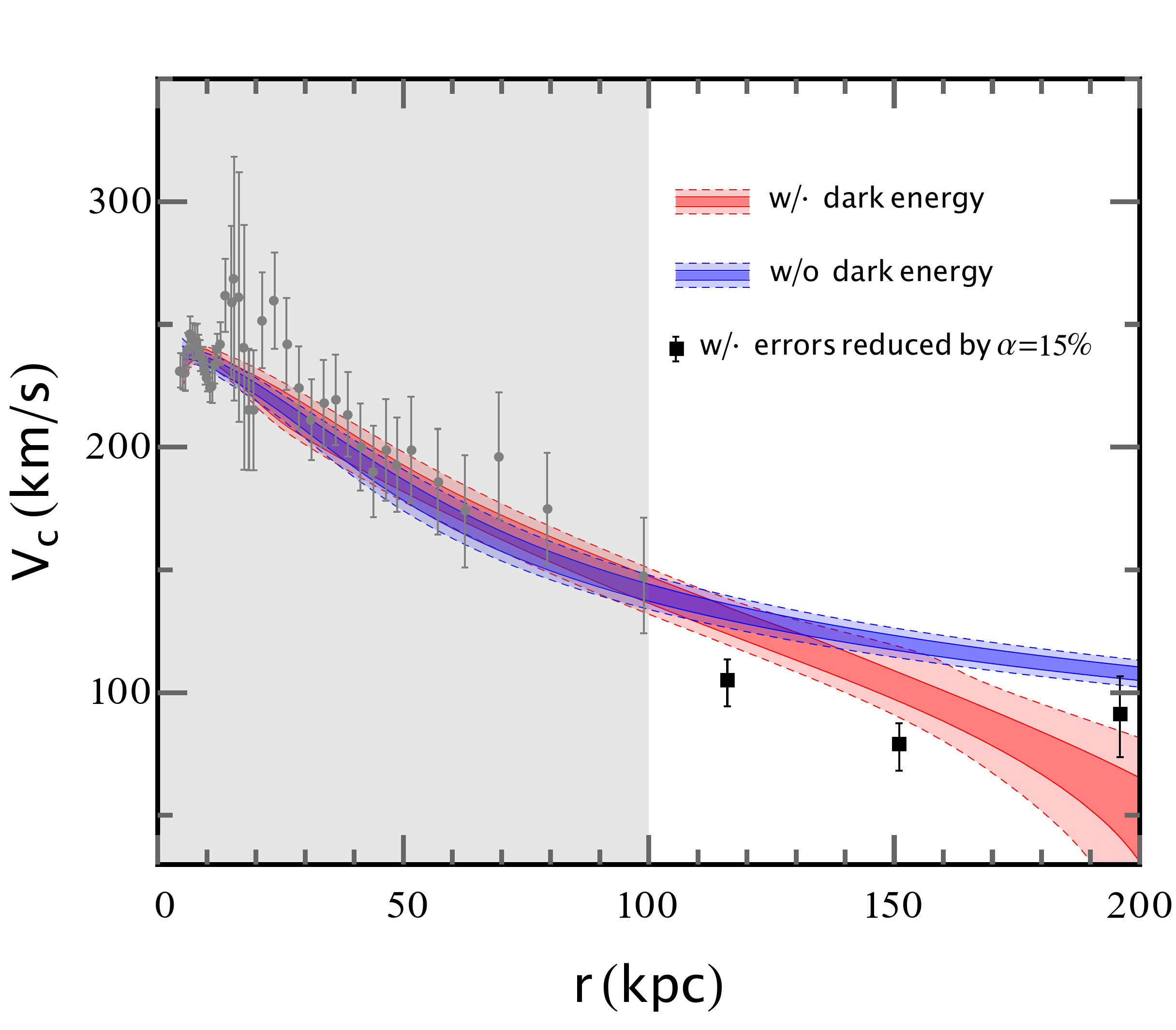}
}
\caption{Fits to the MW rotation curve, with and without the DE contribution included in the fitted models, respectively. In either case, the solid lines represent 1$\sigma$ bounds, and the dashed lines 2$\sigma$ bounds. In the fitting procedure, the errors of RVs at $r\sim100-200~\mbox{kpc}$ are ideally assumed to be $\alpha=15\%$ times their presently measured ones. }
\label{fig:emcee}
\end{figure}

For comparison, we also perform a fit to the MW rotation curve only using the galactic mass model.
In this case, the contribution from the dark force is completely neglected. 
Figure~\ref{fig:emcee} shows the difference between the two cases; the rotation curve in the case with the contribution of the dark force included in the fitted model tends to drop faster than that without including the contribution of the dark force into the fitted model, especially in the outer MW region with $r\sim100-200$~kpc.
When $\alpha=15\%$, the rotation curves in the two cases will show a significant difference at the $\sim2.4\sigma$ confidence level in the outer MW region. 
As demonstrated in Figure~\ref{fig:emcee}, the difference clearly implies that the anomalous drop of the rotation curve in the outer MW region can be explained as the dark force effect. 
This difference become rather significant at large distances, i.e. larger than 150~kpc. 
So we need to do more high-accuracy measurements in the outer MW region to provide robust evidence for DE in the future.

For a better understanding of the uncertainty in the rotation curve fitting, we perform a separate fit using only the RV data within the range of $r=4.5$ to $100~\mbox{kpc}$. 
The corresponding result is already included in Table~\ref{tab:fit}. 
As the table shows, $w\sim-0.79^{+0.01}_{-0.02}$. 
At the 1$\sigma$ confidence level, this value differs slightly from the result obtained between $r=4.5$ and $200~\mbox{kpc}$.
However, they coincide with each other at the 2$\sigma$ confidence level.
In both fits, the value of the EoS parameter deviates significantly from $w=-1.028\pm0.031$ that was obtained by fitting the $w$CDM model to the cosmological data~\cite{Aghanim:2018eyx}, although it still agrees with~\cite{Zhao:2017cud} within $\sim1.2\sigma$ errors.
The deviation may be indicative of the existence of DE beyond the CC model,
or it could suggest errors in determining the profile of the MW halo.

\begin{figure}
\centerline{
\includegraphics[width=0.68\columnwidth]{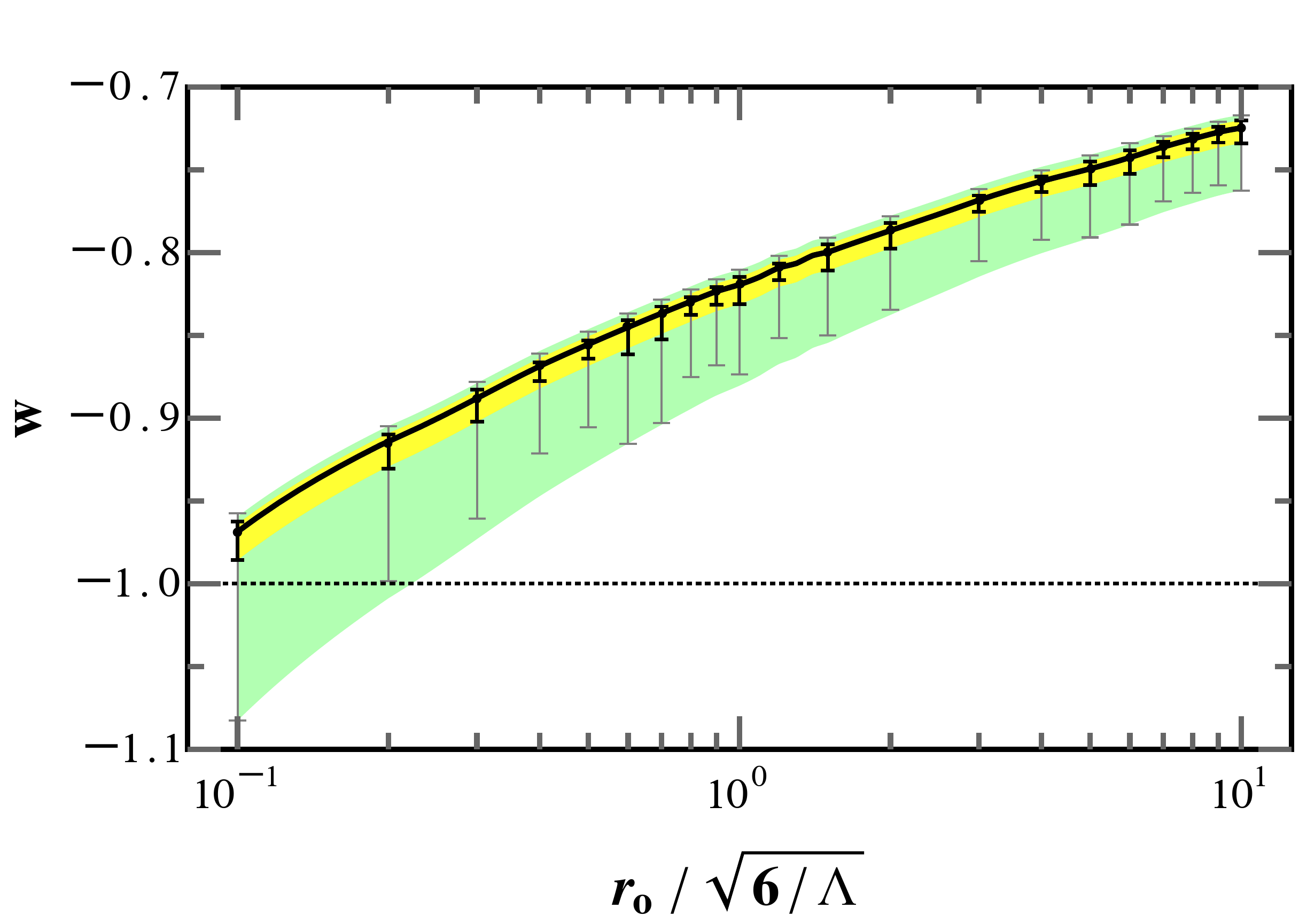}
}
\caption{The best-fit results of $w$ for various $r_{\rm{o}}$ values. The black points correspond to the best-fit values, and the 1$\sigma$ and 2$\sigma$ error bars are depicted in black and gray, respectively. In particular, $w=-1$ is denoted by the black dotted line.} 
\label{fig:w2sigma.vs.ro}
\end{figure}

We also investigate the impact of the chosen $r_{\rm{o}}$ value on the results. Specifically, we perform fittings on the MW data in the range of $r=4.5$ to $200~\mbox{kpc}$ using different $r_{\rm{o}}$ values and show in Figure~\ref{fig:w2sigma.vs.ro} the resulting values of $w$ as a function of $r_{\rm{o}}$ (see Table~\ref{tab:dependence} for details). As illustrated in the figure, when $r_{\rm{o}}\gtrsim\,0.1\,\sqrt{6/\Lambda}$, the fitted value of $w$ is always greater than -1, and it increases as $r_{\rm{o}}$ becomes larger. Therefore, it can be expected that $w\approx-1$ occur in the range where $r_{\rm{o}}<0.1\,\sqrt{6/\Lambda}$. However, in this range, reliable error estimates for the fitted values of $w$ can not be obtained. To facilitate the understanding of the findings, we introduce a parameter $\Lambda_{w\,\approx-1}\sim6 / r_{\rm{o}}^2$ when $w$ is nearly equal to -1~\cite{Zhang:2021ygh}, with $\frac{\Lambda_{w\,\approx-1}}{8\pi}$ denoting the energy density of cosmological DE with $w\approx-1$. If the fitted values of $w$ are consistent with $w=-1$, we can use $\frac{\Lambda_{w\,\approx-1}}{8\pi}$ to estimate the energy density of DE in cosmology. In the range of $r_{\rm{o}}<0.1\,\sqrt{6/\Lambda}$, the value of $\Lambda_{w\,\approx-1}$ is much greater than $\Lambda=4.24\times10^{-66}~\mbox{eV}^2$~\cite{Aghanim:2018eyx}. Exactly speaking, it indicates that the energy density of DE can be at least 2 orders of magnitude higher than the currently measured value. If not, the obtained values of $w$ will be larger than that of $w=-1$. In either case, the fitting results obtained in the regime of $r_{\rm{o}}<0.1\,\sqrt{6/\Lambda}$ could be a hint for the existence of DE beyond the CC model, provided that we have a good understanding of the dark matter halo in the MW galaxy.

\section{Conclusions}\label{discussion}

The mysterious DE poses a great challenge to modern science. 
So far various efforts have been devoted to explore the DE's origin and nature.
In this work, we compared the DE's role in an expanding universe and that in a gravitationally self-bounded system, investigated the dark force induced by DE,
and made our attempts at determining its EoS parameter $w$ on astrophysical scales through measuring the effects of the dark force on rotation curves.

First, we extended the standard RW metric to a more general case by introducing the ERW metric. 
Then, based on the Einstein equation, we derived the generalized Friedmann equation in the ERW spacetime, 
with its curvature term expressed as the average of the sectional curvatures, as well as presented the generalized conservation equations,
and found from these basic equations in cosmology that the cosmological evolution is determined by the dynamical pressure, 
where the dynamical pressure is defined as negative one-third the trace of the stress tensor components, rather than one of the component.

Second, we further investigated the SdS$_{w}$ spacetime and explored the origin and nature of the dark force.
Under gravitational fields, DE may redistribute itself, especially when $w\neq-1$, 
otherwise it will violate the constraint from the Einstein equation. 
However, in this general case, the energy-momentum tensor for DE can still take the standard Weinberg’s isotropic form, 
which is form-invariant under general spatial rotations. 
In a curved spacetime, the EoS parameter $w$ cannot be defined in the same way as that of the traditional perfect fluid in fluid mechanism.
Nevertheless it can be defined as the ratio between the dynamical pressure and the energy density, which is compatible with current theories and cosmological observations.
By the definition, we established a connection in the EoS parameter between DE in the universe and its counterpart in the SdS$_{w}$ spacetime,
laying a theoretical foundation for exploring the origin and nature of the dark force.

Third, we investigated the dark force in analogy to what we have done in Newtonian gravity.
By solving the Einstein equation, we demonstrated that the gravitational potential induced by DE with a generic EoS parameter $w$ still satisfies the Poisson equation.
Then, according to the additivity and linearity, we derived a model-independent form for the total gravitational potential including both the matter and DE contributions.
Thus, the analytical form of the dark force can be obtained by taking the gradient of the DE potential.
Accordingly, we further studied the repulsion of the dark force, 
and found that, for galaxies, the RV values of any objects bound to them can be affected significantly by the dark force at large galactocentric distances;
for instance, the observed RV values should be smaller than those predicted by the purely Newtonian force.

Then, by the repulsion of the dark force, we newly proposed a method to detect DE on astrophysical scales, 
independently of cosmology and specific DE models.
By fitting rotation curves, we can constrain the DE parameter $w$ and obtain accurate values for different $r_{\rm{o}}$ values, 
despite the large errors in observed RV values beyond 100 kpc from the galactic center. 
As a result, we found the deviations of DE from the CC model, unless some significant mistakes were made in understanding the halo profile of the dark matter in the MW galaxy.

Finally, we discussed the prospects for using our method to detect DE in the MW galaxy. 
As more RV measurements take place at large galactocentric radii, the DE parameter $w$ will be measured to a higher accuracy.
When the accuracy is increased to a certain degree, the best-fit rotation curve which includes the dark force contribution will deviate from that does not include the dark force contribution, especially in the outer MW region with $r\sim100-200$~kpc. 
For example, once the uncertainties of RVs in the outer MW region decrease to $\sim15\%$ of the present-day ones, the deviation can be confirmed at $\sim2.4\sigma$ confidence level. It clearly indicates that future prospects for the DE detection through the effects of the dark force on astrophysical scales are quite bright.


\section{Acknowledgements}
The authors contributed equally. We thank Prof. Q.-H. Cao and Prof. S.-L. Xiong for their support as well as the anonymous reviewers for their valuable comments and constructive suggestions. This work is partially supported by the National Program on Key Research and Development Project (Grant No. 2021YFA0718500) from the Minister of Science and Technology of China. RZ acknowledges the support by the National Natural Science Foundation of China (Grant Nos. 12075257, 12235001, 12273042) as well as the funding from the Institute of High Energy Physics, Chinese Academy of Sciences (Grant No. Y6515580U1) and the funding from Chinese Academy of Sciences (Grant No. Y8291120K2). ZZ acknowledges the support by the Institute of High Energy Physics (Grant No. E25155U1) and the support by the Strategic Priority Research Program on Space Science of the Chinese Academy of Sciences (Grant No. XDA15052700). 
\bibliography{draft}
\bibliographystyle{JHEP}


\appendix

\newpage

\section{Conservation of energy and momentum}
\label{App:A}

Generally, a spacetime metic can be written as
\begin{eqnarray}
\label{eq:metric}
\mathrm{d} s^2=g_{\mu\nu}\,\mathrm{d} x^{\mu}\mathrm{d} x^{\nu},
\end{eqnarray}
with $\mu,\nu=0,1,2,3$, where $g_{\mu\nu}$ is the covariant component of the metric tensor. 
As shown in equation~\eqref{eq:ERW}, the nonzero components of the ERW metric and the inverse metric are:
\begin{eqnarray}
\label{eq:ERWmetricC}
g_{00}&\!=\!&Z^2\left(x,y,z\right)\equiv constant,~~ g_{ii}=-a^2\left(t\right)\,R^2\left(x,y,z\right),~~~\\
g^{00}&\!=\!&\frac{1}{Z^2\left(x,y,z\right)}\equiv constant,~~ g^{ii}=-\frac{1}{a^{2}\left(t\right)R^{2}\left(x,y,z\right)}.
\end{eqnarray}
From these we can get the Christoffel symbols, given by 
\begin{eqnarray}
\label{eq:Christoffel}
\Gamma^{\lambda}_{\,\,\,\mu\nu}=\frac{1}{2}\,g^{\lambda\rho}\left(\partial_{\mu}\,g_{\rho\nu}+\partial_{\nu}\,g_{\rho\mu}-\partial_{\rho}\,g_{\mu\nu}\right).
\end{eqnarray}
Exactly, they are
\begin{eqnarray}
\label{eq:CgristoffelSs}
\nonumber
\Gamma^0_{\,\,\,ii}&=&-\frac{1}{2}\,g^{00}\,\partial_0\left(g_{ii}\right),\quad \Gamma^i_{\,\,\,0i}=+\frac{1}{2}\,g^{ii}\,\partial_0\left(g_{ii}\right)=\Gamma^i_{\,\,\,i0}\,, \nonumber\\
\nonumber
\Gamma^i_{\,\,\,ii}&=&+\frac{1}{2}\,g^{ii}\,\partial_i\left(g_{ii}\right),~\quad \Gamma^i_{\,\,\,jj}=-\frac{1}{2}\,g^{ii}\,\partial_i\left(g_{jj}\right)\,, \nonumber\\
\Gamma^i_{\,\,\,ij}&=&+\frac{1}{2}\,g^{ii}\,\partial_j\left(g_{ii}\right)=\Gamma^i_{\,\,\,ji}. 
\end{eqnarray}

As a consequence of conservation of energy and momentum, $T^{\mu\nu}$ satisfies the exact conservation equation:
\begin{align}
\label{eq:S1:EPConservation}
0=\nabla_{\nu}\,T^{\nu\mu}=\partial_{\nu}\,T^{\nu\mu}+\Gamma^{\nu}_{\,\,\,\nu\rho}\,T^{\rho\mu}+\Gamma^{\mu}_{\,\,\,\nu\rho}\,T^{\nu\rho}.
\end{align}
For $\mu=0$, we therefore have
\begin{align}\label{eq:S1:EConservation0}
0=&\partial_\nu T^{0\nu}+\Gamma^0_{\,\,\,\nu\rho}T^{\rho\nu}+\Gamma^\nu_{\,\,\,\nu0}T^{00}\nonumber\\
=&\partial_0 T^{00}+\partial_i T^{0i}-\frac{1}{2}\partial^0(g_{ii})T^{ii}+\frac{1}{2}g^{ii}\partial_0(g_{ii})T^{00}\nonumber\\
=&\partial_0 (g^{00}T^{0}_{\,\,\,0})+\partial_i (g^{00}T^{i}_0)-\frac{1}{2}g^{00}\partial_0(g_{ii})g^{ii}T^{i}_{\,\,\,i}+\frac{1}{2}g^{ii}\partial_0(g_{ii})g^{00}T^{0}_{\,\,\,0}\nonumber\\
=&g^{00}\bigg[\partial_0 T^{0}_{\,\,\,0}+\frac{1}{g^{00}}\partial_i (g^{00}T^{i}_{\,\,\,0})-\frac{1}{2}\partial_0(g_{ii})g^{ii}T^{i}_{\,\,\,i}+\frac{1}{2}g^{ii}\partial_0(g_{ii})T^{0}_{\,\,\,0}\bigg],
\end{align}
where $i=1,2,3$. Substituting equations~\eqref{eq:ERWmetricC} and~\eqref{eq:CgristoffelSs} into the last line gives
\begin{align}\label{eq:S1:EConservation1}
0=&\frac{\mathrm{d} T^{0}_{\,\,\,0}}{\mathrm{d} t}+\frac{1}{g^{00}}\,\partial_i\left(g^{00}T^{i}_{\,\,\,0}\right)-\frac{1}{2\left(R^2a^2\right)}\frac{\mathrm{d} \left(R^2a^2\right)}{\mathrm{d} t}T^{i}_{\,\,\,i}+\frac{3}{2\left(R^2a^2\right)}\frac{\mathrm{d} \left(R^2a^2\right)}{\mathrm{d} t}T^{0}_{\,\,\,0}\nonumber\\
=&\frac{\mathrm{d} T^{0}_{\,\,\,0}}{\mathrm{d} t}+\frac{3\dot{a}}{a}\left(T^0_{\,\,\,0}-\frac{1}{3}T^{i}_{\,\,\,i}\right)+\frac{1}{g^{00}}\partial_i \left(g^{00}T^{i}_{\,\,\,0}\right)\nonumber\\
=&\frac{\mathrm{d} T^{0}_{\,\,\,0}}{\mathrm{d} t}+3\,\frac{\dot{a}}{a}\left(T^0_{\,\,\,0}-\frac{1}{3}T^{i}_{\,\,\,i}\right)+\frac{2\partial_i Z}{Z}T^{i}_{\,\,\,0}+\partial_i T^{i}_{\,\,\,0}. 
\end{align}
For $\mu=i$, we obtain
\begin{align}\label{eq:S1:PConservation0}
0=&\partial_\nu T^{i\nu}+\Gamma^i_{\,\,\,\nu\rho}T^{\rho\nu}+\Gamma^\nu_{\,\,\,\nu\rho}T^{i\rho}\nonumber\\
=&(\partial_0 T^{i0}+\partial_j T^{ij})+(2\Gamma^i_{\,\,\,i0}T^{0i}+\Gamma^i_{\,\,\,ii}T^{ii}+\Gamma^i_{\,\,\,ll}T^{ll}+\Gamma^i_{\,\,\,il}T^{li})+(\Gamma^j_{\,\,\,j0}T^{i0}+\Gamma^k_{\,\,\,kj}T^{ij})\nonumber\\
=&\partial_j T^{ij}+\Gamma^i_{\,\,\,ii}T^{ii}+\Gamma^i_{\,\,\,ll}T^{ll}+\Gamma^i_{\,\,\,il}T^{li}+\Gamma^k_{\,\,\,kj}T^{ij}\nonumber\\
&+\partial_0 T^{i0}+\Gamma^j_{\,\,\,j0}T^{i0}+2\Gamma^i_{\,\,\,i0}T^{0i}\nonumber\\
=&\partial_j T^{ij}+\frac{1}{2}g^{ii}(\partial_ig_{ii})T^{ii}
-\frac{1}{2}g^{ii}(\partial_ig_{ll})T^{ll}+\frac{1}{2}g^{ii}(\partial_lg_{ii})T^{il}
+\frac{1}{2}g^{kk}(\partial_jg_{kk})T^{ij}\nonumber\\
&+\partial_0 T^{i0}+\frac{1}{2}g^{jj}(\partial_0g_{jj})T^{i0}+g^{ii}(\partial_0g_{ii})T^{0i}\nonumber\\
=&\partial_j T^{ij}+\frac{1}{2}g^{ii}(\partial_ig_{ii})T^{ii}
-\frac{1}{2}g^{ii}(\partial_ig_{ii})T^{jj}+\frac{1}{2}g^{ii}(\partial_jg_{ii})T^{ij}
+\frac{3}{2}g^{ii}(\partial_jg_{ii})T^{ij}\nonumber\\
&+\partial_0 T^{i0}+\frac{3}{2}g^{ii}(\partial_0g_{ii})T^{0i}+g^{ii}(\partial_0g_{ii})T^{0i}\nonumber\\
=&\partial_j T^{ij}+\frac{1}{2}g^{ii}(\partial_ig_{ii})(T^{ii}-T^{jj})+2g^{ii}(\partial_jg_{ii})T^{ij}\nonumber\\
&+\partial_0 T^{i0}+\frac{5}{2}g^{ii}(\partial_0g_{ii})T^{0i}\nonumber\\
=&\partial_j (g^{ii}T^{i}_{\,\,\,j})+\frac{1}{2}(g^{ii})^2(\partial_ig_{ii})(T^{i}_{\,\,\,i}-T^{j}_{\,\,\,j})+2(g^{ii})^2(\partial_jg_{ii})T^{i}_{\,\,\,j}\nonumber\\
&+\partial_0 (g^{ii}T^{0}_{\,\,\,i})+\frac{5}{2}(g^{ii})^2(\partial_0g_{ii})T^{0}_{\,\,\,i}\nonumber\\
=&g^{ii}\Big[\partial_j T^{i}_{\,\,\,j}+g_{ii}\partial_j (g^{ii})T^{i}_{\,\,\,j}+\frac{1}{2}g^{ii}(\partial_ig_{ii})(T^{i}_{\,\,\,i}-T^{j}_{\,\,\,j})
+2g^{ii}(\partial_jg_{ii})T^{i}_{\,\,\,j}\nonumber\\
&+\partial_0 T^{0}_{\,\,\,i}+g_{ii}\partial_0 (g^{ii})T^{0}_{\,\,\,i}+\frac{5}{2}g^{ii}(\partial_0g_{ii})T^{0}_{\,\,\,i}\Big]\nonumber\\
=&g^{ii}\Big[\partial_j T^{i}_{\,\,\,j}+\frac{1}{2}g^{ii}(\partial_ig_{ii})(T^{i}_{\,\,\,i}-T^{j}_{\,\,\,j})+g^{ii}(\partial_jg_{ii})T^{i}_{\,\,\,j}\nonumber\\
&+\partial_0 T^{0}_{\,\,\,i}+\frac{3}{2}\,g^{ii}(\partial_0g_{ii})T^{0}_{\,\,\,i}\Big],
\end{align}
where $i,j,k\in\{1,2,3\}$ and $l\in\{1,2,3\}-\{i\}$. Combining equations~\eqref{eq:ERWmetricC} and~\eqref{eq:CgristoffelSs} with this equation~\eqref{eq:S1:PConservation0} yields
\begin{align}\label{eq:S1:PConservation1}
0=&\partial_j T^{i}_{\,\,\,j}+\frac{1}{2}g^{ii}(\partial_ig_{ii})(T^{i}_{\,\,\,i}-T^{j}_{\,\,\,j})+g^{ii}(\partial_jg_{ii})T^{i}_{\,\,\,j}
+\partial_t T^{0}_{\,\,\,i}+\frac{3}{2}g^{ii}(\partial_0g_{ii})T^{0}_{\,\,\,i}\nonumber\\
=&\partial_j T^{i}_{\,\,\,j}+\frac{1}{R}\frac{\mathrm{d} R}{\mathrm{d} x^i}(T^{i}_{\,\,\,i}-T^{j}_{\,\,\,j})+\frac{2}{R}\frac{\mathrm{d} R}{\mathrm{d} x^j}T^{i}_{\,\,\,j}
+\partial_t T^{0}_{\,\,\,i}+3\,\frac{\dot{a}}{a}T^{0}_{\,\,\,i}.
\end{align}

At any point $P$ in curved spacetime, there is a locally inertial frame around the point $P$, comoving with the fluid element. 
As seen by the comoving observer in this inertial frame, the $0i$ and $i0$ components of the energy-momentum tensor satisfy $T_{0i}=T_{i0}=0$ \cite{Maggiore:2018}; more analysis details are presented in appendix~\ref{App:G}. Thus, $T^{0}_{\,\,\,i}=T^{i}_{\,\,\,0}=0$. We therefore obtain
\begin{align}
\label{eq:S1:EConservation2}
0=&\frac{\mathrm{d} T^{0}_{\,\,\,0}}{\mathrm{d} t}+3\,\frac{\dot{a}}{a}\left(T^0_{\,\,\,0}-\frac{1}{3}T^{i}_{\,\,\,i}\right),\\
\label{eq:S1:PConservation2}
0=&\partial_j T^{i}_{\,\,\,j}+\frac{1}{R}\frac{\mathrm{d} R}{\mathrm{d} x^i}(T^{i}_{\,\,\,i}-T^{j}_{\,\,\,j})+\frac{2}{R}\frac{\mathrm{d} R}{\mathrm{d} x^j}T^{i}_{\,\,\,j},
\end{align}
directly from equations~\eqref{eq:S1:EConservation1} and~\eqref{eq:S1:PConservation1}, respectively. In general relativity, the former expresses conservation of energy in the ERW universe, while the later corresponds to conservation of the $i$th component of the momentum.

In the standard RW spacetime, the stress tensor of a perfect fluid is given by
\begin{align*}
T^{i}_{\,\,\,j}=\hat{\pi}\,\delta^{i}_{\,\,\,j}.
\end{align*} 
Plugging this expression into equation\,\eqref{eq:S1:PConservation2} gives
\begin{align*}
\partial_i\,\hat{\pi}=0,
\end{align*}
which is the well-known conservation equation that corresponds to conservation of momentum in the standard RW cosmology (see appendix~\ref{App:D} for more details).

\section{The Friedmann equations}
\label{App:B}

Here we present the steps of deriving the Friedmann equations from the ERW metric.
The first step we would take is to calculate the Ricci tensor from the formula
\begin{align*}
R_{\mu\sigma}&=\Gamma^\nu_{\,\,\,\mu\sigma,\nu}-\Gamma^\nu_{\,\,\,\sigma\nu,\mu}+\Gamma^\lambda_{\,\,\,\mu\sigma}\Gamma^\nu_{\,\,\,\lambda\nu}-\Gamma^\nu_{\,\,\,\mu\lambda}\Gamma^\lambda_{\,\,\,\nu\sigma}.
\end{align*}
Then, for $\left(\mu,\nu\right)=\left(0,0\right)$, one gets
\begin{align*}
R_{00}=&\Gamma^\nu_{\,\,\,00,\nu}-\Gamma^\nu_{\,\,\,0\nu,0}+\Gamma^\lambda_{\,\,\,00}\Gamma^\nu_{\,\,\,\lambda\nu}-\Gamma^\nu_{\,\,\,0\lambda}\Gamma^\lambda_{\,\,\,\nu0}\\
=&-\frac{1}{2}\partial_0(g^{ii}\partial_0g_{ii})-\frac{1}{4}(g^{ii}\partial_0g_{ii})^2\\
=&-3\,\frac{\ddot{a}}{a},
\end{align*}
while for $\left(\mu,\nu\right)=\left(i,i\right)$, one has
\begin{align*}
R_{ii}=&\Gamma^\nu_{\,\,\,ii,\nu}-\Gamma^\nu_{\,\,\,i\nu,i}+\Gamma^\lambda_{\,\,\,ii}\Gamma^\nu_{\,\,\,\lambda\nu}-\Gamma^\lambda_{\,\,\,i\nu}\Gamma^\nu_{\,\,\,\lambda i}\\
=&-\frac{1}{2}\partial_0(g^{00}\partial_0 g_{ii})+\frac{1}{2}\partial_i(g^{ii}\partial_i g_{ii})-\frac{1}{2}\partial_\ell(g^{\ell\ell}\partial_\ell g_{ii})\\
&-\frac{1}{2}\partial_i(g^{jj}\partial_ig_{jj})\\%
&-\frac{1}{4}g^{00}g^{jj}(\partial_0 g_{ii})(\partial_0 g_{jj})+\frac{1}{4}g^{ii}g^{jj}(\partial_ig_{ii})(\partial_ig_{jj})
-\frac{1}{4}g^{\ell\ell}g^{kk}(\partial_\ell g_{ii})(\partial_\ell g_{kk})\\
&+\frac{1}{2}g^{00}g^{ii}(\partial_0g_{ii})^2-\frac{1}{4}(g^{jj}\partial_ig_{jj})^2
+\frac{1}{2}g^{ii}g^{\ell\ell}(\partial_\ell g_{ii})^2\\
=&-\frac{1}{2}\partial_0(g^{00}\partial_0 g_{ii})+\partial_i(g^{ii}\partial_i g_{ii})-\frac{1}{2}\partial_j(g^{jj}\partial_j g_{ii})\\
&-\frac{1}{2}\partial_i(g^{jj}\partial_ig_{jj})\\
&-\frac{1}{4}g^{00}g^{jj}(\partial_0 g_{ii})(\partial_0 g_{jj})+\frac{1}{2}g^{ii}g^{jj}(\partial_ig_{ii})(\partial_ig_{jj})
-\frac{1}{4}g^{jj}g^{kk}(\partial_jg_{ii})(\partial_jg_{kk})\\
&+\frac{1}{2}g^{00}g^{ii}(\partial_0g_{ii})^2-\frac{1}{4}(g^{jj}\partial_ig_{jj})^2
-\frac{1}{2}(g^{ii})^2(\partial_ig_{ii})^2+\frac{1}{2}g^{ii}g^{jj}(\partial_jg_{ii})^2\\
=&-\frac{1}{2}g^{00}\partial^2_0 g_{ii}+(\partial_ig^{ii})(\partial_i g_{ii})+g^{ii}\partial^2_i g_{ii}
-\frac{1}{2}g^{ii}\partial^2_j g_{ii}-\frac{1}{2}(\partial_jg^{ii})(\partial_j g_{ii})\\
&-\frac{3}{2}g^{ii}\partial^2_i g_{ii}-\frac{3}{2}(\partial_ig^{ii})(\partial_ig_{ii})\\
&-\frac{3}{4}g^{00}g^{ii}(\partial_0 g_{ii})(\partial_0 g_{ii})+\frac{3}{2}(g^{ii})^2(\partial_ig_{ii})(\partial_ig_{ii})
-\frac{3}{4}(g^{ii})^2(\partial_jg_{ii})(\partial_jg_{ii})\\
&+\frac{1}{2}g^{00}g^{ii}(\partial_0g_{ii})^2-\frac{3}{4}(g^{ii}\partial_ig_{ii})^2
-\frac{1}{2}(g^{ii})^2(\partial_ig_{ii})^2+\frac{1}{2}(g^{ii})^2(\partial_jg_{ii})^2\\
=&-\frac{1}{2}g^{00}\partial^2_0 g_{ii}-\frac{1}{4}g^{00}g^{ii}(\partial_0g_{ii})^2\\
&-\frac{1}{2}(\partial_ig^{ii})(\partial_ig_{ii})-\frac{1}{2}g^{ii}\partial^2_i g_{ii}
-\frac{1}{2}(\partial_jg^{ii})(\partial_j g_{ii})-\frac{1}{2}g^{ii}\partial^2_j g_{ii}\\%
&+\frac{1}{4}(g^{ii})^2(\partial_ig_{ii})^2-\frac{1}{4}(g^{ii})^2(\partial_jg_{ii})^2\\[1.5mm]
=&-g^{00}g_{ii}\bigg[\frac{\ddot{a}}{a}+2\left(\frac{\dot{a}}{a}\right)^2\bigg]-\frac{\partial^2_i R}{R}-\frac{\partial^2_j R}{R}+2\,\left(\frac{\partial_i R}{R}\right)^2, 
\end{align*}
where $i,j,k\in\{1,2,3\}$ and $l\in\{1,2,3\}-\{i\}$. Accordingly, one writes the Ricci scalar as 
\begin{align*}
R=& g^{00}R_{00}+g^{ii}R_{ii}\\
=&-3g^{00}\frac{\ddot{a}}{a}-3g^{00}\bigg[\frac{\ddot{a}}{a}+2\left(\frac{\dot{a}}{a}\right)^2\bigg]-g^{ii}\frac{\partial^2_i R}{R}-g^{ii}\frac{\partial^2_j R}{R}+2g^{ii}(\frac{\partial_i R}{R})^2\\[1.5mm]
=&-6\,g^{00}\bigg[\frac{\ddot{a}}{a}+\left(\frac{\dot{a}}{a}\right)^2\bigg]-4\,g^{ii}\frac{\partial^2_i R}{R}+2\,g^{ii}(\frac{\partial_i R}{R})^2,
\end{align*}
where $i,j\in\{1,2,3\}$. 

We now turn to the Einstein equation. Recall that it can be written in the form:
\begin{align*}
G^{\nu}_{\,\,\,\mu}=R^{\nu}_{\,\,\,\mu}-\frac{1}{2}\,\delta^{\nu}_{\,\,\,\mu}\,R=T^{\nu}_{\,\,\,\mu}.
\end{align*}
The $\mu\nu=00$ equation gives 
\begin{align}
\nonumber
8\pi \,T^{0}_{\,\,\,0}=&R^{0}_{\,\,\,0}-\frac{1}{2}R\\
\nonumber
=&3\,g^{00}\left(\frac{\dot{a}}{a}\right)^2+2\,g^{ii}\frac{\partial^2_i R}{R}-g^{ii}\left(\frac{\partial_i R}{R}\right)^2\\
\label{eq:Frie2d0}
=&3\,g^{00}\left(\frac{\dot{a}}{a}\right)^2-\frac{2}{a^2 R^2}\frac{\partial^2_i R}{R}+\frac{1}{a^2 R^2}\left(\frac{\partial_i R}{R}\right)^2,
\end{align}
and the $\mu\nu=ii$ equations is
\begin{align}\label{eq:Frie1d0}
8\pi \,T^{i}_{\,\,\,i}=&R^{i}_{\,\,\,i}-\frac{1}{2}R\nonumber\\
=&g^{00}\bigg[2\frac{\ddot{a}}{a}+\left(\frac{\dot{a}}{a}\right)^2\bigg]+g^{ll}\frac{\partial^2_l R}{R}-g^{ll}\left(\frac{\partial_l R}{R}\right)^2+g^{ii}\left(\frac{\partial_i R}{R}\right)^2\nonumber\\
=&g^{00}\bigg[2\frac{\ddot{a}}{a}+\left(\frac{\dot{a}}{a}\right)^2\bigg]-\frac{1}{a^2R^2}\frac{\partial^2_l R}{R}+\frac{1}{a^2R^2}\left(\frac{\partial_l R}{R}\right)^2-\frac{1}{a^2R^2}\left(\frac{\partial_i R}{R}\right)^2,
\end{align}
where $i,j,k\in\{1,2,3\}$ and $l\in\{1,2,3\}-\{i\}$. 
Using~\eqref{eq:Frie2d0} to eliminate the first derivative in~\eqref{eq:Frie1d0}, we can obtain the generalized Friedmann equations:
\begin{align}
\label{eq:Friedmann2g}
g^{00}\,\frac{\ddot{a}}{a}=&-\frac{4}{3}\pi \left(T^{0}_{\,\,\,0}-T^{i}_{\,\,\,i}\right),\\[1.5mm]
\label{eq:Friedmann1g}
g^{00}\,\left(\frac{\dot{a}}{a}\right)^2=&\frac{8}{3}\pi T^{0}_{\,\,\,0}-\frac{K\left(x,y,z\right)}{a^2},
\end{align}
with (see appendix~\ref{App:C} for more details)
\begin{align*}
K\left(x,y,z\right)=\frac{1}{3}\,\bigg[-\frac{2}{R^2}\frac{\partial^2_i R}{R}+\frac{1}{R^2}\left(\frac{\partial_i R}{R}\right)^2\bigg],
\end{align*}
where $i,j,k\in\{1,2,3\}$ and $l\in\{1,2,3\}-\{i\}$.
As we will show bellow, $K=K\left(x,y,z\right)$ is intrinsically an effective sectional curvature. 
When $Z^2\left(x,y,z\right)\equiv1$, it reduces to
\begin{align}
\label{eq:Friedmann2}
\frac{\ddot{a}}{a}=&-\frac{4}{3}\pi \left(T^{0}_{\,\,\,0}-T^{i}_{\,\,\,i}\right),\\[1.5mm]
\label{eq:Friedmann1}
\left(\frac{\dot{a}}{a}\right)^2=&-\frac{K\left(x,y,z\right)}{a^2}+\frac{8}{3}\pi \,T^{0}_{\,\,\,0},
\end{align}
which reduce to the standard Friedmann equations when $K$ is a constant.

\section{Sectional curvature}
\label{App:C}

Denote $\mathrm{d} \sigma^{2}$ as the spatial part of the ERW metric~\eqref{eq:ERW} with $a=1$, namely
\begin{eqnarray}
\label{eq:SpaceMetric}
\mathrm{d} \sigma^{2}=\gamma_{ij}\,\mathrm{d} x^{i} \mathrm{d} x^{j}=R\left(x,y,z\right)^{2}\left(\mathrm{d} x^{\,2}\!+\!\mathrm{d} y^{\,2}\!+\!\mathrm{d} z^{\,2}\right).
\end{eqnarray}
whose fully covariant version of the curvature tensor of type $\left(0,4\right)$ is given by
\begin{eqnarray}
\label{eq:SMetric}
^{\left(3\right)}\!\!R_{ijkm}=\gamma_{is}\,^{\left(3\right)}\!\!R^{s}_{jkm}=\gamma_{is}\,\bigg[{}^{\left(3\right)}\!\Gamma^s_{\,\,\,jm,k}-{}^{\left(3\right)}\!\Gamma^s_{\,\,\,jk,m}+{}^{\left(3\right)}\!\Gamma^p_{\,\,\,jm}{}^{\left(3\right)}\!\Gamma^s_{\,\,\,pk}-{}^{\left(3\right)}\!\Gamma^p_{\,\,\,jk}{}^{\left(3\right)}\!\Gamma^s_{\,\,\,pm}\bigg],
\end{eqnarray}
where $i,j,k,m\in\{1,2,3\}$. Here we use the superscript $^{\left(3\right)}$ to indicate that it is associated with the 3-metric~\eqref{eq:SpaceMetric}.

Then the {\it sectional curvature} of a given surface at point $p$ can be described by
\begin{eqnarray}
\label{eq:SMetric}
\nonumber
K_{p}=K_{p}[i,j,k,m]=-\frac{^{\left(3\right)}\!\!R_{ijkl}}{\gamma_{ik}\gamma_{jm}-\gamma_{im}\gamma_{jk}},
\end{eqnarray}
which is also named as the Gaussian curvature\cite{Zhang:2021ygh}.
It has a clear meaning. For instance, in the special case of the standard RW spacetime~\eqref{eq:RWMetric3}, one has
\begin{eqnarray}
\label{eq:RWSpaceMetric}
\mathrm{d} \sigma^{2}&=&-\frac{\mathrm{d} x^2+\mathrm{d} y^2+\mathrm{d} z^2}{\big[1+\frac{1}{4}\kappa \left(x^{2}+y^{2}+z^{2}\right)\big]^2}.
\end{eqnarray}
where $\kappa$ is a constant parameter.
In this case, for $\forall~i,j,k,m\in\{1,2,3\}$, $K_{p}=\kappa={\rm Constant}$ (see appendix~\ref{App:D} for the derivations). It means that the 3-space defined by~\eqref{eq:RWSpaceMetric} is actually a constant curvature space. 

Our task is clear. We need to understand the ERW spacetime~\eqref{eq:ERW}. Especially, we need to fully understand the $K-$term in the generalized Friedmann equations. 
In the general case of the ERW spacetime, direct calculation reveals that
\begin{eqnarray}
\label{eq:Kterm}
\nonumber
K_{p}^{i}&=&K_{p}[j,k,j,k]\\[1.5mm]
\nonumber
&=&-\frac{1}{R^2}\bigg[\left(\frac{\partial_j R}{R}\right)^2-\frac{\partial^2_j R}{R}\!+\!\left(\frac{\partial_k R}{R}\right)^2-\frac{\partial^2_k R}{R}-\left(\frac{\partial_i R}{R}\right)^2\!\bigg],
\end{eqnarray}
where $i\in\{1,2,3\}$, $j\in\{1,2,3\}-\{i\}$ and $k\in\{1,2,3\}-\{i,j\}$.
Define
\begin{eqnarray}
\label{eq:Kterm}
\nonumber
K\left(x,y,z\right)=\frac{~\sum K^i_{p}~}{~3~}.
\end{eqnarray}
Then we have 
\begin{eqnarray}
\label{eq:ERW-Kterm}
\nonumber
K\left(x,y,z\right)=\frac{1}{3}\,\bigg[-\frac{2}{R^2}\frac{\partial^2_i R}{R}+\frac{1}{R^2}\left(\frac{\partial_i R}{R}\right)^2\bigg],
\end{eqnarray}
which is obviously a generalization of the constant curvature $k$ in equation~\eqref{eq:RWSpaceMetric}.

\section{The Robertson-Walker spacetime}
\label{App:D}

A specific example of the ERW metric is the standard RW metric. It can be written in the following form,
\begin{eqnarray}
\label{eq:RWMetric}
\mathrm{d} s^{2}&=&\mathrm{d} t^2-a^2\left(t\right)\bigg[\frac{\mathrm{d} \chi^2}{1-\kappa\, \chi^2}+\chi^2 \mathrm{d} \Omega^2\bigg],
\end{eqnarray}
where $\mathrm{d} \Omega^2=\mathrm{d} \theta^2\!+\!\sin^2\theta\,\mathrm{d} \varphi^2$ is the metric of the unit 2-sphere, and $\chi$ is a radial coordinate.

A second form of the standard RW metric is obtained from~\eqref{eq:RWMetric} via the relation
\begin{align}
\chi=\frac{r}{1+\frac{1}{4}\kappa r^{2}},
\end{align}
namely:
\begin{eqnarray}
\label{eq:RWMetric2}
\mathrm{d} s^{2}&=&\mathrm{d} t^2-\bigg[a\left(t\right)R\left(r\right)\bigg]^2\left(\mathrm{d} r^2+r^2 \mathrm{d} \Omega^2\right),
\end{eqnarray}
with 
\begin{eqnarray}
\label{eq:Rfactor}
R\left(r\right)=\frac{1}{1+\frac{1}{4}\kappa r^{2}},
\end{eqnarray}
where $\kappa$ is a real number.
In the same coordinate system as the ERW metric, the standard RW metric can be expressed as
\begin{eqnarray}
\label{eq:RWMetric3}
\mathrm{d} s^{2}&=&\mathrm{d} t^2-\bigg[a\left(t\right)R\left(r\right)\bigg]^2\left(\mathrm{d} x^2+\mathrm{d} y^2+\mathrm{d} z^2\right),
\end{eqnarray}
with $r^2=x^2+y^2+z^2$. Its spatial part with $a=1$ then reads 
\begin{eqnarray}
\label{eq:RWspaceMetric3}
\mathrm{d} \sigma^{2}&=&-\frac{\mathrm{d} x^2+\mathrm{d} y^2+\mathrm{d} z^2}{\big[1+\frac{1}{4}\kappa \left(x^{2}+y^{2}+z^{2}\right)\big]^2}.
\end{eqnarray}
Thus, the space with this metric form is a constant curvature space, and its sectional curvature is $\kappa$.

\section{Methodologies} 
\label{App:E}

Now we provide further clarifications on the literature bellow, making it clear
that our methodology is still independent and self-contained.
(i). In the RW spacetime, the cosmological expansion effect of DE was incorporated in a modified Poisson equation \cite{Balaguera-Antolinez:2007csw}: 
\begin{eqnarray}
\label{MPa1}
\nabla^2 \Phi=4\pi\,\delta\rho_\mathrm{m}-3\,\frac{\ddot{a}}{a},
\end{eqnarray}
where $\delta\rho_\mathrm{m}$ is the density fluctuation of matter around its cosmological background $\bar{\rho}_\mathrm{m}=\bar{\rho}_\mathrm{m}(a)$,
and $\frac{\ddot{a}}{a}$ is a function of the cosmological expansion factor $a=a(t)$.
Generally, one has
\begin{eqnarray}
\label{MPa2}
\Phi=\Phi_\mathrm{m}-\frac{\ddot{a}}{a}\,r^2=-\int\frac{\delta\rho_\mathrm{m}(\vec{r^\prime})}{\mid \vec{r}-\vec{r^\prime}\mid}\mathrm{d} \vec{r^\prime}-\frac{\ddot{a}}{a}\,r^2, 
\end{eqnarray}
where the matter potential $\Phi_\mathrm{m}$ is solely determined by $\delta\rho_\mathrm{m}$.
In a real astrophysical system, $\frac{\ddot{a}}{a}$ can be treated as a constant.
So the second term in this equation is quite different from the DE term shown in equation~\eqref{eq:totalPotential}.
In fact, equations~\eqref{MPa1} and \eqref{MPa2} are only applicable on a much larger scale than that of a galaxy.
This can be confirmed by reviewing the assumptions and approximations presented in \cite{Balaguera-Antolinez:2007csw}. 
For instance, the matter density can be decomposed into a background value plus a perturbation,
with the matter background $\bar{\rho}_\mathrm{m}$ being completely determined by the expansion factor, as assumed in \cite{Balaguera-Antolinez:2007csw}. 
On astrophysical scales, such as that of the MW galaxy, the assumption breaks down as the large structures, such as the galaxy's dark matter halo, cannot be treated as mere perturbations.
Additionally, the cosmological evolution of the expansion factor $a$ highly relies on the components other than DE.
In this methodology, the DE effect is indirectly included through the expansion factor, making it inherently dependent on cosmology.
Therefore, this methodology is only appropriate for studying DE on astrophysical scales only within the context of a specific cosmological model,
which differs significantly from the one used in this work. 
(ii). In the literature~\cite{Balaguera-Antolinez:2006nwo,Ho:2015nsa}, they introduced the Poisson equation with the cosmological constant $\Lambda$: 
\begin{eqnarray}
\label{CCa1}
\nabla^2 \Phi=4\pi(\rho_\mathrm{m}+3\,p_\mathrm{m})-\Lambda,
\end{eqnarray}
where $\rho_\mathrm{m}$ and $p_\mathrm{m}$ are presented in the main text.
Compared with our generalized Poisson equation~\eqref{PoEtotal}, this equation is just a special case.
Accordingly, we have
\begin{eqnarray}
\label{CCa2}
\Phi=\Phi_\mathrm{m}-\frac{1}{6}\Lambda \,r^2,
\end{eqnarray}
which agrees with that shown in equation~\eqref{eq:totalPotential} for $w=-1$.
Clearly, it can well describe the astrophysical object surrounded by the CC dark energy. 
However, in order to derive a general potential containing DE with a generic EoS parameter, i.e., $w\not\equiv-1$, for a realistic astrophysical system, 
we have to overcome various difficulties in advance.
These include expressing the energy-momentum tensor of DE in a general isotropic form, 
defining the EoS parameter physically in a curved spacetime, 
and associating the isotropic energy-momentum tensor with a DE model using the physical quantities such as the dynamical pressure and the EoS parameter (see section~\ref{ThA} for details).
For instance, as shown by equation~\eqref{eq:Tcomponets}, the energy-momentum tensor has already been expressed in the general isotropic form,
in which its off-diagonal components vanish only if $w=-1$, such as in the CC case. 
However, in the general case of $w\neq-1$, the isotropic energy-momentum tensor has not been fully described or analyzed in the literature;
one of the difficulties in this scenario is the emergence of the non-zero off-diagonal components that depend on the choice of coordinate system.
The generalized Poisson equation, which includes contributions from both matter and DE, can only be derived after solving those problems. 
Once the equation is derived, we can present the exact form of the dark force in a realistic astrophysical system by calculating the gradient of the potential. 
If these problems are not resolved, the methodology presented in this work cannot be applied to investigate DE on astrophysical scales.
So we have adopted an unusual methodology to propose a cosmology-independent method for detecting various forms of DE on astrophysical scales, 
without relying on any specific DE models.
This methodology forms the basis of our approach and enables us to detect DE independently of any cosmological assumptions or DE models.

\section{Supplementary data}
\label{SM}
\label{App:F}

We vary the value of $r_{\mbox{o}}$ to investigate its effect on the $w$ values obtained from fitting the MW data between $r=4.5$ and $200~\mbox{kpc}$. Table~\ref{tab:dependence} displays the resulting value of $w$ for each $r_{\mbox{o}}$ value. It is evident from the table that as $r_{\mbox{o}}$ becomes larger, the fitting value of $w$ increases, whereas the other parameters such as the ones for the disc and dark halo remain almost unchanged. Specifically, one has $r_{\rm{\scriptsize d}}\approx 2.9~{\rm(kpc)}$, $\rho_{\mbox{\scriptsize h,0}}\approx0.010 ~M_{\odot}\,{\rm pc}^{-3}$, and $r_{\mbox{\scriptsize h}}\approx~ 18{\rm kpc}$ for the disk and halo.

\begin{table}
\centerline{
\begin{tabular}{|c|cccc|c|}
\hline
$r_{\mbox{o}}$ ($\sqrt{6/\Lambda}$)&$r_{\rm{\scriptsize d}} {\rm(kpc)}$&$\rho_{\mbox{\scriptsize h,0}}$ ($M_{\odot}\,{\rm pc}^{-3}$)&$r_{\mbox{\scriptsize h}} {\rm(kpc)}$&$w$&$\chi^2_{\mbox{\scriptsize red}}$\\
\hline
0.1&$3.0_{-0.1}^{+0.2}$&$0.012_{-0.003}^{+0.002}$&$16_{-3}^{+1}$&$-0.969_{-0.017}^{+0.006}$&$0.87$\\
\hline
0.2&$2.9_{-0.1}^{+0.2}$&$0.012_{-0.003}^{+0.002}$&$17_{-3}^{+1}$&$-0.915_{-0.015}^{+0.006}$&$0.86$\\
\hline
0.3&$2.9_{-0.1}^{+0.2}$&$0.011_{-0.003}^{+0.002}$&$17_{-3}^{+1}$&$-0.888_{-0.014}^{+0.005}$&$0.86$\\
\hline
0.4&$2.9_{-0.1}^{+0.1}$&$0.011_{-0.002}^{+0.002}$&$17_{-2}^{+1}$&$-0.868_{-0.009}^{+0.002}$&$0.85$\\
\hline
0.5&$2.9_{-0.1}^{+0.1}$&$0.011_{-0.002}^{+0.001}$&$17_{-2}^{+1}$&$-0.855_{-0.009}^{+0.002}$&$0.86$\\
\hline
0.6&$2.9_{-0.1}^{+0.2}$&$0.011_{-0.003}^{+0.002}$&$18_{-6}^{+8}$&$-0.845_{-0.016}^{+0.004}$&$0.86$\\
\hline
0.7&$2.9_{-0.1}^{+0.2}$&$0.011_{-0.003}^{+0.001}$&$18_{-3}^{+1}$&$-0.837_{-0.016}^{+0.004}$&$0.86$\\
\hline
0.8&$2.9_{-0.1}^{+0.1}$&$0.011_{-0.002}^{+0.001}$&$18_{-2}^{+1}$&$-0.830_{-0.008}^{+0.002}$&$0.84$\\
\hline
0.9&$2.9_{-0.1}^{+0.1}$&$0.011_{-0.002}^{+0.001}$&$18_{-2}^{+1}$&$-0.823_{-0.008}^{+0.002}$&$0.84$\\
\hline
1&$2.9_{-0.1}^{+0.2}$&$0.011_{-0.003}^{+0.002}$&$18_{-3}^{+1}$&$-0.819_{-0.012}^{+0.005}$&$0.85$\\
\hline
2&$2.9_{-0.2}^{+0.2}$&$0.010_{-0.003}^{+0.002}$&$18_{-3}^{+1}$&$-0.787_{-0.011}^{+0.004}$&$0.85$\\
\hline
3&$2.9_{-0.1}^{+0.1}$&$0.010_{-0.002}^{+0.001}$&$18_{-2}^{+1}$&$-0.768_{-0.007}^{+0.003}$&$0.84$\\
\hline
4&$2.9_{-0.1}^{+0.1}$&$0.010_{-0.002}^{+0.001}$&$19_{-2}^{+1}$&$-0.757_{-0.007}^{+0.003}$&$0.83$\\
\hline
5&$2.9_{-0.1}^{+0.2}$&$0.010_{-0.003}^{+0.002}$&$19_{-3}^{+2}$&$-0.749_{-0.010}^{+0.004}$&$0.84$\\
\hline
6&$2.9_{-0.1}^{+0.2}$&$0.010_{-0.003}^{+0.001}$&$19_{-3}^{+1}$&$-0.742_{-0.010}^{+0.004}$&$0.84$\\
\hline
7&$2.8_{-0.1}^{+0.1}$&$0.010_{-0.002}^{+0.001}$&$19_{-2}^{+1}$&$-0.736_{-0.006}^{+0.003}$&$0.83$\\
\hline
8&$2.9_{-0.1}^{+0.2}$&$0.011_{-0.003}^{+0.002}$&$18_{-3}^{+1}$&$-0.731_{-0.006}^{+0.003}$&$0.83$\\
\hline
9&$2.9_{-0.1}^{+0.1}$&$0.010_{-0.002}^{+0.001}$&$19_{-2}^{+1}$&$-0.727_{-0.006}^{+0.003}$&$0.83$\\
\hline
10&$2.9_{-0.1}^{+0.2}$&$0.010_{-0.002}^{+0.001}$&$19_{-3}^{+2}$&$-0.725_{-0.009}^{+0.005}$&$0.84$\\
\hline
\end{tabular}
}
\caption{Best-fit parameters for various $r_{\mbox{o}}$ values, obtained from the MW data over the range $r=4.5$ to $200~\mbox{kpc}$.
This is very similar to Table~\ref{tab:fit}. 
The errors are within 1$\sigma$ confidence level. We also show the reduced $\chi^2\,$ in the last column. }
\label{tab:dependence}
\end{table}

\newpage

\section{Further comments}
\label{App:G}

Based on the Hubble law, the recession velocity $\upsilon$ at time $t$ can be expressed as~\cite{Carroll2014,Rindler2006book} 
\begin{eqnarray}
\label{eq:HLaw}
\upsilon=\dot{d}_{\rm P}=H\,d_{\rm P},
\end{eqnarray}
where $H=\frac{\dot{a}}{a}$ is defined as the Hubble parameter exclusively as a function of time $t$, and $d_{\rm P}$ is the instantaneous physical distance. 
The linearity of the Hubble law means that all galaxies recede from the observer at time $t$ at velocities linearly proportional to their distances from the observer.
According to the linearity, the metric \eqref{eq:ERW} can always be reparametrized so that $Z\left(x,y,z\right)\equiv constant$. 
To prove this, let us define $\eta=\eta\left(t,x,y,z\right)$ in such a way that 
\begin{eqnarray}
\label{eq:Z0eta}
Z_{0}\,\mathrm{d}\eta=Z\left(x,y,z\right)\mathrm{d}t,
\end{eqnarray}
where $\eta$ corresponds to the proper time, and $Z_{0}$ is a global constant. Thus, $a\left(t\right)=a\left(\eta,x,y,z\right)$.
If the Hubble law holds, there exist functions $a^{\ast}=a^{\ast}\left(\eta\right)$ and $R^{\ast}=R^{\ast}\left(x,y,z\right)$ 
such that 
\begin{eqnarray}
\label{eq:a.vs.Rast}
\big[a\left(\eta,x,y,z\right)R\left(x,y,z\right)\big]^{2}\left(\mathrm{d} x^{\,2}+\mathrm{d} y^{\,2}+\mathrm{d} z^{\,2}\right)=\big[a^{\ast}\left(\eta\right)R^{\ast}\left(x_{\ast},y_{\ast},z_{\ast}\right)\big]^{2}\left(\mathrm{d} x_{\ast}^{\,2}+\mathrm{d} y_{\ast}^{\,2}+\mathrm{d} z_{\ast}^{\,2}\right),~~~~~~
\end{eqnarray}
which is required by the linearity of the Hubble law. Here, the coordinates $\left(x_{\ast},y_{\ast},z_{\ast}\right)$ are reparametrized space-like ones. Accordingly, the metric becomes
\begin{eqnarray}
\label{eq:ERW-app.G}
\begin{array}{rcl}
\displaystyle
\mathrm{d} s^{2}=Z_{0}^{2}\,\mathrm{d} \eta^{2}-\big[a^{\ast}\left(\eta\right)R^{\ast}\left(x_{\ast},y_{\ast},z_{\ast}\right)\big]^{2}\left(\mathrm{d} x_{\ast}^{\,2}+\mathrm{d} y_{\ast}^{\,2}+\mathrm{d} z_{\ast}^{\,2}\right).
\end{array}
\end{eqnarray}
If the left-hand side of equation \eqref{eq:a.vs.Rast} cannot be expressed as a product of a function that depends only on time and a function that depends only on spatial variables,
$a^{\ast}$ will be no longer solely dependent on time $\eta$. 
Instead, it may also depend on the spatial coordinates, i.e., $a^{\ast}=a^{\ast}\left(\eta,x_{\ast},y_{\ast},z_{\ast}\right)$. 
In this case, the Hubble parameter has to be defined as 
\begin{eqnarray}
\label{eq:Hpara2}
H=\frac{\dot{a}^{\ast}}{a^{\ast}}=\frac{\partial a^{\ast}\left(\eta,x_{\ast},y_{\ast},z_{\ast}\right)\!/\partial\eta}{a^{\ast}\left(\eta,x_{\ast},y_{\ast},z_{\ast}\right)},
\end{eqnarray}
which is space-dependent. Actually, it violates the Hubble law. Specifically, it is inconsistent with the linearity of the Hubble law.
If this equation were true, distant space regions would expand faster than nearby ones, and thus the evolution of the universe would be spatially unstable.
On the other hand, as seen by the comoving observer, $T_{0i}=T_{i0}=0$. 
Correspondingly, in the comoving frame of the observer we now have $G_{0i}=G_{i0}=0$.
Therefore, from the metric \eqref{eq:ERW}, we obtain 
\begin{eqnarray}
\label{eq:G0iGi0=0}
G_{0i}=G_{i0}=4\,\frac{\dot{a}}{\,a\,}\,\frac{\partial_i Z}{\,Z\,}=0.
\end{eqnarray}
Then, we immediately find $Z\left(x,y,z\right)\equiv constant$. 
In fact, $Z\left(x,y,z\right)$ can only be reduced to a global constant if a comoving system of coordinates is chosen.

\end{document}